\documentclass[a4paper,12pt]{article}

\usepackage{amsmath, amsthm, natbib} 
\usepackage{tikz-network}
\usepackage{graphicx}
\usepackage{xcolor}
\usepackage{subcaption}  
\usepackage{newtxtext,newtxmath} 
\usepackage[scaled=0.5]{helvet} 
\usepackage{inconsolata}         
\usepackage{setspace}
\usepackage[utf8]{inputenc}   
\usepackage[T1]{fontenc}      

\usepackage{geometry}
\newtheorem{prop}{Proposition}

\newtheorem{cor}{Corollary}

\newtheorem*{claim}{Claim}
\theoremstyle{definition}

\newtheorem{example}{Example}

\newcommand{\mR}{\mathbb{R}}
\newcommand{\mE}{\mathbb{E}}
\newcommand{\ms}{\mathbf{s}}

\newcommand{\mone}{\mathbf{1}}

\newcommand{\mc}{\mathbf{c}}

\newcommand{\mb}{\mathbf{b}}

\newcommand{\tx}{\tau_{x}}
\newcommand{\ty}{\tau_{y}}

\newcommand{\ga}{\gamma}

\usepackage[hidelinks]{hyperref}

\usepackage[title]{appendix}

\title{Title of the Paper}
\author{Author Name\thanks{Affiliation, Email}}
\date{\today}

\newcommand{\JEL}[1]{\vspace{0.1cm}\par\noindent {\small{\textbf{JEL Classification}:} #1}}

\begin{document}
\title{Network Heterogeneity and Value of Information}
\author{Kota Murayama\thanks{Email: KotaMurayama@cuhk.edu.hk. This paper is based on a chapter of my dissertation. I am grateful to my advisors, Asher Wolinsky, Alessandro Pavan, and Jeffrey Ely, for their continuous encouragement and guidance. I also thank  Mariagiovanna Baccara, Piotr Dworczak, Yingni Guo, Bruno Strulovici, Harry Pei, Pinchuan Ong, Toomas Hinnosaar, Takashi Ui, Faruk Gul, Arjada Bardi, Eddie Dekel, Matthew Jackson, Ben Golub, Tomasz Strzalecki, Hideshi Itoh, Kohei Kawamura, Daisuke Hirata, Shintaro Miura, and seminar and conference participants. The author would like to acknowledge the financial support of the Hong Kong Research Grants Council under the Early Career Scheme 24621822.} 
\\The Chinese University of Hong Kong}

\maketitle
\begin{abstract}
Does greater connectivity enhance the value of public information? I study a networked beauty contest game where agents balance adaptation to the fundamental with local coordination. The analysis reveals a stark non-monotonicity: while public disclosure improves welfare when interactions are uniform, regardless of their intensity, it can be detrimental in core-periphery structures. This welfare loss stems from a distortion driven by the core, where core agents over-respond to a noisy public signal, forcing peripheral neighbors to absorb this volatility to maintain alignment. These findings suggest that standard transparency policies can backfire in tiered markets where dominant hubs propagate excess volatility.
\vspace{1em} 
\\ \noindent \textbf{Keywords:} Beauty contest, coordination, value of information, network games. 
 \vspace{0.5em} 
 \JEL{C72, D83, D85}
  \end{abstract}

\newpage
\section{Introduction}
Many economic interactions exhibit local strategic complementarities, where agents seek to coordinate with a specific subset of partners rather than with the entire population. In production networks, for instance, a firm's optimal pricing depends on the pricing decisions of its direct suppliers and distributors. Similarly, in over-the-counter (OTC) financial markets, dealers' willingness to hold inventory depends on their ability to rebalance positions through their specific inter-dealer connections.

While coordination motives are defined by a local neighborhood, agents' beliefs are often anchored by public information. Central bank inflation forecasts shape firms' macroeconomic expectations, which in turn inform their pricing decisions \citep{CoibionEtAl2018, Coibion2020-yr}. In OTC markets, regulators utilize public stress tests to stabilize lending, providing a system-wide signal that reduces opacity and coordinates market beliefs \citep{Petrella2013, MorganEtAl2014, Flannery2017}.

Standard intuition suggests that public disclosures are most effective when coordination incentives are uniformly strong, as public signals serve as a focal point for global alignment. Where interactions are weak or sparse, the social value of public information is therefore expected to diminish. This paper revisits this intuition by asking: How does the shape of interaction networks alter the welfare implications of public information?

To address this question, I study a beauty contest game embedded in a general interaction network. Agents face two competing objectives: an adaptation motive to match the fundamental state and a coordination motive to align their actions with those of their neighbors. The interaction network defines both the intensity and direction of coordination motives. This framework captures the key trade-offs in the motivating examples: firms adjusting prices to aggregate costs or demand while aligning with specific partners, or dealers setting quotes to track asset values while maintaining spread relative to counterparties.

The analysis reveals a stark non-monotonicity between the value of public information and network connectivity. While public information is beneficial when interactions are uniformly weak or uniformly strong, it can be detrimental in the intermediate case where interactions are concentrated among a few hub agents.
I show that providing noisy public information harms welfare when the network exhibits sharp structural heterogeneity, specifically in core-periphery networks characterized by small, dense cores. 
This result arises from a core-driven distortion. Core agents, prioritizing internal coordination, anchor heavily on the public signal and deviate from the fundamental state.
This deviation exacerbates the strategic trade-off for peripheral agents: as the core's actions and the fundamental state diverge, peripheral agents are induced to sacrifice alignment with the fundamental to maintain coordination with the core. When the core is small and dense, the welfare loss arising from this intensified trade-off dominates the direct informational benefits of the signal.

These topological distinctions map directly to real-world market structures. While hub-and-spoke architectures with sparse inter-hub connectivity characterize production networks \citep[e.g.,][]{Atalay2011, Mizuno2014, Bernard2019}, tiered architectures with dense cores are typical of OTC markets \citep[e.g.,][]{CraigVonPeter2014, Hollifield2017, LiSchurhoff2019}. Since the welfare impact of public disclosure depends on core density, a regulator operating without precise knowledge of the underlying network risks implementing counterproductive policies. My findings thus underscore the potential for substantial welfare gains from acquiring granular network data.

More formally, I formulate the problem within a linear-quadratic-Gaussian framework, which admits a unique linear equilibrium. In this equilibrium, an agent's responsiveness to public information is proportional to her Katz-Bonacich centrality. This centrality measure aggregates an agent's total connectivity by summing all discounted paths of any length emanating from her. Intuitively, these paths represent the agent's indirect coordination motives. For instance, a path to agent $k$ through neighbor $j$ captures $i$'s derived incentive to align with $k$ to coordinate with $j$. These indirect motives amplify the equilibrium weight on the public signal because it serves as the common focal point for such higher-order alignments. Thus, agents with higher centrality place greater weight on the public signal.

The first main result establishes that providing noisy public information reduces an agent's equilibrium payoff if and only if her Katz-Bonacich centrality is sufficiently low relative to the aggregate centrality of her neighbors. To illustrate, consider a simple three-agent network comprising Ann, Bob, and Carol. Bob and Carol are reciprocally connected, forming a dense core, while Ann connects solely to Bob. Reflecting his high centrality, Bob excessively responds to the public signal, rendering his action noisy relative to the fundamental. Ann, in her peripheral position, is forced into a strategic trade-off: to maintain coordination with Bob, she must absorb this induced volatility into her own payoff. When the centrality gap between Ann and Bob is large, the cost of this imported volatility outweighs the direct informational benefit of the signal, resulting in a net payoff loss for Ann.

These distributional effects translate into a precise condition for an aggregate welfare loss. 
I show that the provision of noisy public information becomes socially detrimental if and only if the network features agents who have a combination of disproportionally high Katz-Bonacich centrality and high in-degree. This condition decomposes the mechanism of the welfare loss: high centrality determines the magnitude of the distortion (i.e., how much the agent over-reacts to public information), while high in-degree determines the scale of its transmission (i.e., how many others intend to follow her action). High centrality alone is insufficient to generate aggregate loss; the distorted agent must also be influential enough to propagate this volatility to a broad set of neighbors.

Building upon this insight, I also identify a novel source of inefficiency in endogenous information sharing. In an extension allowing for voluntary disclosure, I show that agents may rationally withhold information to avoid this core-driven distortion. A peripheral agent, foreseeing that releasing her signal will induce highly central neighbors to anchor aggressively on it, may choose to suppress the signal to avoid importing the resulting volatility. This equilibrium is often socially inefficient because the agent fails to internalize the coordination benefits the signal would provide to the core, leading to an under-provision of public information.

Finally, while the finding that public information can reduce welfare parallels the seminal result of \cite{Morris2002-pp}, the underlying economic mechanism is distinct. \cite{Morris2002-pp} rely on a preference misalignment where the social planner cares solely about alignment with the fundamental. In contrast, the present result assumes a utilitarian planner, and the welfare loss arises only when there is significant structural heterogeneity in the interaction network, which is absent in their symmetric framework.

\section{Literature}
The literature on the social value of information, pioneered by \cite{Morris2002-pp}, establishes that public information can reduce welfare when private and social incentives diverge (see \cite{Pavan2015-si} and \cite{Angeletos2016-hb} for overviews). Analyzing general symmetric linear-quadratic games, \cite{Angeletos2007-uq} formalize this insight, attributing the value of information to the discrepancy between equilibrium and socially optimal degrees of coordination. \cite{Ui2015-if} subsequently provides a full characterization of the value of information for this class of games. Extensions have identified specific frictions that complicate this baseline, such as dynamic learning externalities \citep{Amador2010-jk} or the crowding out of private information acquisition \citep{Colombo2014-mn}.

Despite these advances, the existing literature predominantly assumes ex-ante payoff symmetry.\footnote{Exceptions include \cite{Huo2020-sv} on dynamic games and \cite{Miyashita2023-dl} on information design, though their focus differs from the analysis here.} I depart from this benchmark by analyzing a general framework of heterogeneous interaction structures. My analysis reveals that such heterogeneity creates a pronounced structural wedge between private and social incentives, substantial enough to reverse the sign of the welfare effect even under a utilitarian planner. This implies that the key determinants of the value of information lie not in global preference parameters, but in the granular network structures through which agents interact.

This paper also relates to the literature on endogenous information acquisition and sharing. The seminal models of endogenous information acquisition in beauty contest games identify strategic complementarities: the marginal value of a signal increases with the number of other agents who also observe it \citep{Hellwig2009-op, Myatt2012-cu}. Recent contributions, such as \cite{Myatt2019-iu} and \cite{Leister2020-fk}, extend this framework to network games built upon \cite{Ballester2006-hq}. In a related model, \cite{Herskovic2020-vd} demonstrate that coordination motives drive agents to endogenously form core-periphery networks to facilitate information exchange. These studies generally find that connectivity reinforces the incentive for shared information, preserving the intuition that agents prefer to coordinate on common signals.

In contrast, I show that sharp core-periphery structures reverse this logic. In such networks, the negative externality of induced aggregate volatility dominates the incentive to coordinate on common signals. This identifies a novel source of inefficiency: peripheral agents may rationally withhold signals to avoid the volatility caused by public disclosure.

More broadly, this paper contributes to the growing literature on optimal interventions in network games. \cite{Ballester2006-hq} focus on structural interventions, identifying the ``key player'' whose removal maximally impacts aggregate activity. \cite{Galeotti2020-em} study incentive interventions, characterizing optimal adjustments to agents' marginal returns. In the domain of information design, \cite{Miyashita2023-dl} develop a semidefinite programming approach to characterize the optimal information structure in general linear-quadratic-Gaussian games. Unlike these approaches, which typically derive agent-specific optima, I focus on public disclosure, which is the standard policy instrument in decentralized markets where fine-tuned intervention is often infeasible.

\section{Model}
I introduce a beauty contest game in which agents' coordination motives are represented as a directed and weighted network.

\vspace{1em}
\noindent \textbf{Payoff Function.}
There are $n$ agents, indexed by $i \in N\equiv \{1, \ldots, n \}$, where $n \geq 2$.
Agent $i$'s action is $a_{i}\in \mathbb{R}$, and $a = (a_{i})_{i \in N}$ denotes an action profile.
There is a common payoff state $\theta \in \mR$.
Each agent $i$'s payoff function is

\begin{equation}\label{payoff}
u_{i}(a, \theta)= -\left(1-\sum_{j \neq i}g_{ij}\right)(a_{i}-\theta)^{2} - \sum_{j \neq i}g_{ij}(a_{i}-a_{j})^{2}.\end{equation}

The first term in (\ref{payoff}) represents the loss from the deviation of agent $i$'s action from the payoff state. The second term is a sum of losses from deviations of $i$'s action from those of other agents.
The coefficient $g_{ij}\geq 0$ measures the strength of $i$'s coordination motive with $j$.
Throughout the paper, I assume  that $\sum_{j\neq i}g_{ij}<1$ for all $i \in N$. 
An $ n\times n$ matrix $G=[g_{ij}]$ collects these coordination motives, where $g_{ii}=0$ for all $i \in N$.
This matrix $G$ can be viewed as the adjacency matrix of a weighted and directed network and is referred to as an \textit{interaction network}.
I call $d^{out}_i\equiv \sum_{j\neq i} g_{ij}$ $i$'s \textit{out-degree} and call $j$ a \textit{neighbor} of $i$ if $g_{ij}>0$.

\vspace{1em}
\noindent \textbf{Information Structure.}
The information structure is standard, following \cite{Morris2002-pp}.
The payoff state $\theta$ follows an improper uniform prior distribution over the real line.
Each agent $i$ observes a signal vector $\ms_i=(x_i, y)$, where $x_i$ is a private signal
observed only by $i$, and $y$ is a public signal observed by all agents.
Specifically, I define $x_i\equiv \theta + \varepsilon_i$ and $y\equiv \theta + \varepsilon_{y},$ where $\varepsilon_i$ and $\varepsilon_y$ are normally distributed with
$$\mE[\varepsilon_i]=\mE[\varepsilon_y]=0, \mathrm{Var}[\varepsilon_i]=\tau_x^{-1}, \mathrm{Var}[\varepsilon_y]=\tau_y^{-1},$$ and the noise terms $\varepsilon_i, \varepsilon_y$ and $\theta$ are all jointly independent.
The relative precision of private signals is denoted by  $\gamma\equiv \tau_x/(\tau_x+\tau_y) \in (0, 1)$.  
 
\vspace{1em}
\noindent \textbf{Equilibrium.}
The game defined above has a unique linear Bayesian Nash equilibrium.
Since the payoff function (\ref{payoff}) is strictly concave in $a_i$, the first-order condition is necessary and sufficient for equilibrium.
Each agent $i$'s first-order condition is
\begin{equation}\label{FOC}
a_i= \left(1-d_i^{out}\right)\mE_i[\theta]+ \sum_{j \neq i}g_{ij}\mE_i[a_{j}],
\end{equation}
where $\mE_i[\cdot]\equiv\mE[ \cdot \mid \ms_i]$ is the conditional expectation operator, conditional on $i$'s information $\ms_i$.
Consider linear strategies of the form $a_i=b_i y + (1-b_i)x_i$ for all $i \in N$.\footnote{Focusing on this form of linear strategies is without loss of generality. In Section \ref{sectionproof1}, I show that the equilibrium slopes must sum to one for all agents in any linear equilibrium.}
Plugging these linear strategies into (\ref{FOC}) yields
\begin{equation}\label{FOC2}
b_i y + (1-b_i)x_i= \left(1-\ga +\ga \sum_{j \neq i}g_{ij}b_j\right)y+\left(\ga -\ga \sum_{j \neq i}g_{ij}b_j\right)x_i.
\end{equation}
By matching the coefficients in (\ref{FOC2}), the equilibrium slopes must satisfy
\begin{equation}\label{FOC3}
b_i -\ga \sum_{j \neq i}g_{ij}b_j= 1-\ga.
\end{equation}
In matrix notation, these simultaneous equations are written as
\begin{equation}\label{equationequilibrium}
(I - \ga G)\mb =(1-\ga)\mone,\end{equation}
where $I$ is an $n \times n$ identity matrix, and $\mb=(b_1, \ldots, b_n)'$ and $\mone=(1, \ldots, 1)'$ are $n$-dimensional column vectors.
The system of linear equations (\ref{equationequilibrium}) has a unique nonnegative solution for $\mb$ because the inverse matrix $(I-\ga G)^{-1}$ exists and is nonnegative.\footnote{By the Perron-Frobenius theorem, $G$ has the largest nonnegative real eigenvalue $\rho(G)$ and $\rho(G)\leq \max_{i \in N}\sum_{j \neq i}g_{ij}<1$ by assumption.
Also, by Theorem III$^*$ of \cite{Debreu1953-ae}, $(I-\ga G)^{-1}$ exists and is nonnegative if and only if $\ga\rho(G)<1$. Combining these observations proves the claim.}
The vector $\mc(\ga, G)\equiv (I - \ga G)^{-1}\mone$ is called the \textit{Katz-Bonacich centrality vector} in the network literature. 

\begin{prop}\label{propequilibrium}
There exists a unique linear equilibrium in which agent $i$'s strategy is
\begin{equation}\label{equilibrium} a^*_{i}= b^*_i y+ (1-b^*_i) x_i,\end{equation}
where $b^*_i=(1-\ga) c_i(\ga, G)$ is proportional to $i$'s Katz-Bonacich centrality.
\end{prop} 

In equilibrium,  agent $i$'s direct and indirect coordination motives are positively associated with her responsiveness to the public signal.
Intuitively, when $i$'s coordination motive strengthens, the commonality of the public signal becomes more valuable for aligning her action with her neighbors' actions. 
Moreover, if $i$'s neighbors have stronger coordination motives, they will respond more to the public signal, further increasing the value of the public signal as a coordination device.
Thus, not only an agent's own coordination motive but also those of her neighbors (and neighbors' neighbors, etc.) positively affect her responsiveness to the public signal.\footnote{The relative precision of private information, $\ga$, acts as the equilibrium discount factor in the Katz-Bonacich centrality. This role arises from how different orders of agents' expectations about $\theta$ (e.g., $\mE_i[\theta]$ and $\mE_i[\mE_j[\theta]])$ respond to changes in $\ga$. A more detailed intuition is provided in \cite{Murayama2025-ln}.}
Following the standard practice in this literature, I focus on this unique linear equilibrium in the analysis.

\section{Results}\label{sectionpublic}
The main results show that public information provision can be detrimental to an individual agent and aggregate welfare, depending on the geometry of interaction networks. Most proofs and derivations are provided in the appendix. 

\subsection{Individual Payoff}
Let $U_i(\tx, \ty, G)\equiv\mE [u_i(a^*, \theta)]$  denote agent $i$'s  equilibrium  payoff.\footnote{$U_i(\tx, \ty, G)$ is well-defined because $u_i(a^*, \theta)$ is independent of $\theta$.}
To simplify notation, define a random variable $X_i\equiv a_i^*-\theta$ for each $i \in N$.
The variance and covariance of $X_i$ are
\begin{align}\label{identity1}
	\mE[X^2_i]&= \tx^{-1}(1-\ga)c_i(\ga, G)\left(c_i(\ga, G)-2\right)+\tx^{-1};\\
	\label{identity2} \mE[X_iX_j]&=  \ty^{-1}(1-\ga)^2c_i(\ga, G)c_j(\ga, G);\\ 
\sum_{j\neq i}g_{ij}\mE[X_iX_j]&=  \tx^{-1}(1-\ga)c_i(\ga, G)\left(c_i(\ga, G)-1\right).
\label{identity3}
\end{align}
The last equality  (\ref{identity3}) follows from the self-referentiality of the Katz-Bonacich centrality:
\begin{align}
c_i(\ga, G)=1+\ga \sum_{j\neq i}g_{ij} c_{j}(\ga, G).
 \end{align}
Using these expressions, $i$'s equilibrium payoff simplifies to
\begin{align}
	\hspace{-0.5cm}	U_i(\tx, \ty, G)&= -\left(1-d_i^{out}\right)\mE[X^2_i] - \sum_{j \neq i}g_{ij}\mE[(X_i-X_j)^2] \notag \\
&=-\mE[X^2_i]- \sum_{j \neq i}g_{ij}\mE[X_j^2]+2 \sum_{j \neq i}g_{ij}\mE[X_i X_j] \notag \\
&=  \tx^{-1}(1-\ga)\left(c_i^2(\ga, G)-\sum_{j \neq i}g_{ij}c_j(\ga, G)\left(c_j(\ga, G)-2\right)\right)- \tx^{-1}(1+d_i^{out}).\label{equationpayoff}
\end{align}
Let $\Delta U_i(\tx, \ty, G)\equiv U_i(\tx, \ty, G)-U_i(\tx,0, G)$ be the effect of public information provision on agent $i$.\footnote{Abusing notation, I write $U_i(\tx,0, G)=-\tx^{-1}(1+d_i^{out})$ for agent $i$'s equilibrium payoff in the absence of public information.}
The following result is immediate from (\ref{equationpayoff}).
\begin{prop}\label{propindividual}
$\Delta U_i(\tx, \ty, G)<0$ if and only if
\begin{equation}
	c_i^2(\ga, G)<\sum_{j \neq i}g_{ij}c_j(\ga, G)(c_j(\ga, G)-2). \label{inequalityindividual}
\end{equation}
\end{prop}  
Proposition \ref{propindividual} states that the provision of public information is detrimental to agent $i$ if and only if, despite engaging in some coordination, her centrality is sufficiently low relative to an aggregate measure of her neighbors' centralities.
Note that inequality (\ref{inequalityindividual}) never holds when the interaction network is regular (i.e., $d_i^{out}=d$ for all $i \in N$).\footnote{In regular interaction networks, each agent's centrality is given by $1/(1-\ga d)$. Hence, the right-hand side of (\ref{inequalityindividual}) must be smaller than the squared centrality.} 
Thus, at least some heterogeneity in agents' out-degrees is necessary for such a detrimental effect on individual payoff.
The following example illustrates interaction networks that admit these adverse effects.

\begin{figure}[ht!]
    \centering
    \begin{minipage}{0.3\textwidth}
        \centering
        \includegraphics[width=\linewidth]{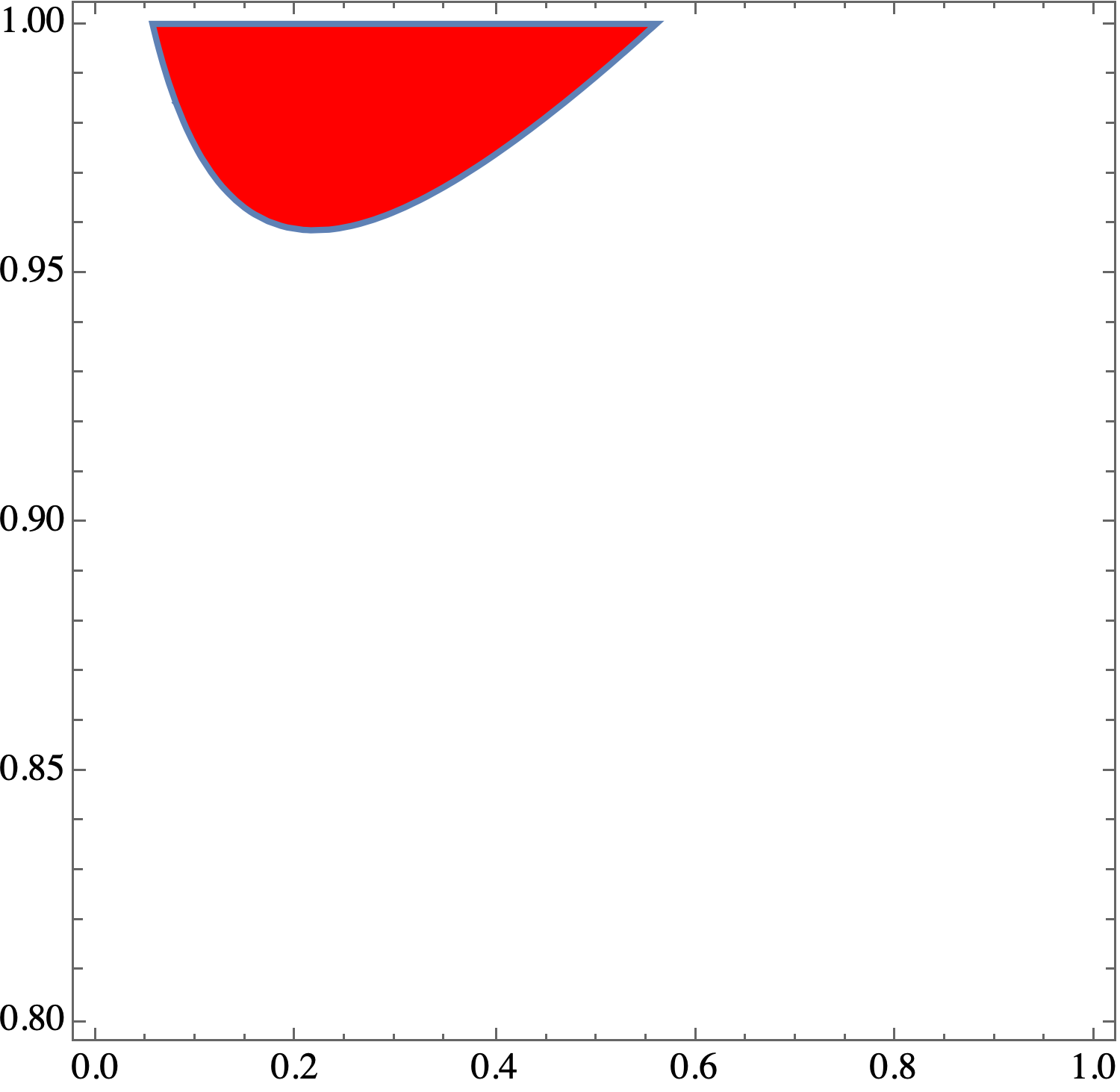}  
        \subcaption{$\gamma = 0.85$}
    \end{minipage}%
    \hspace{0.5cm}
    \begin{minipage}{0.3\textwidth}
        \centering
        \includegraphics[width=\linewidth]{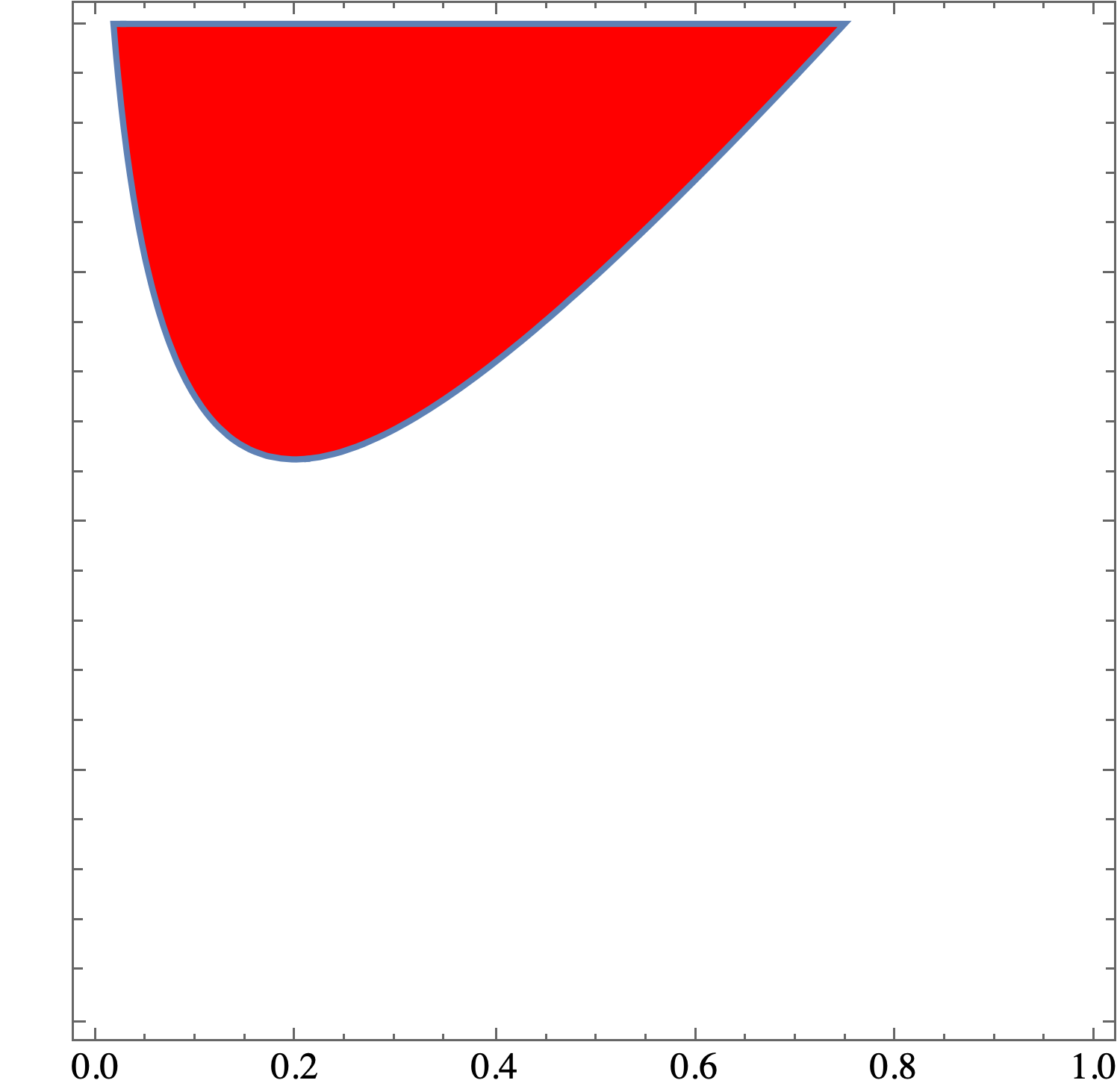}  
        \subcaption{$\gamma = 0.9$}
    \end{minipage}%
      \hspace{0.5cm}
    \begin{minipage}{0.3\textwidth}
        \centering
        \includegraphics[width=\linewidth]{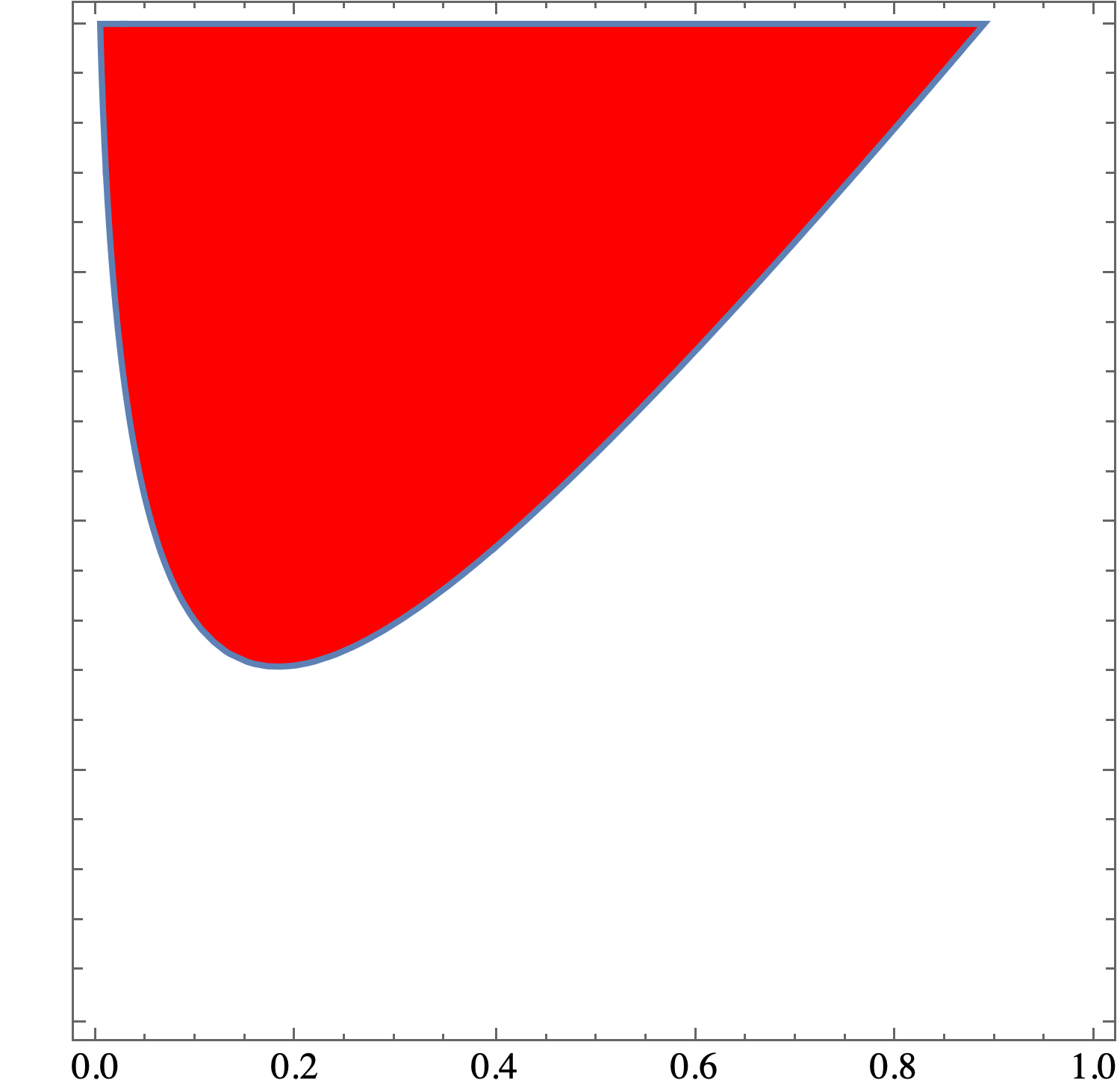}  
        \subcaption{$\gamma = 0.95$}
    \end{minipage} 
        \caption{$\mathcal{G}(\ga)$ for $\gamma\in \{0.85, 0.9, 0.95\}$, where $x$-axis is $\alpha \in [0,1]$ and $y$-axis is $\beta\in [0.8, 1]$.}
    \label{figureindividual}
\end{figure}

\begin{example}\label{exampleABC}
There are three agents, Ann, Bob, and Carol, in a team.	
Bob and Carol know each other well and intend to coordinate their action with each other. Ann is new to the team and intends to follow Bob's action, while Bob and Carol do not intend to follow Ann's action.
Formally, the interaction network is given by $g_{AB}=\alpha$, $g_{BC}=g_{CB}=\beta$, and $g_{AC}=g_{BA}=g_{CA}=0$ (see Figure \ref{figureABC}).
\begin{figure}[ht]\centering
\begin{tikzpicture}[scale=0.9, every node/.style={scale=0.9}]
    \node[draw, circle, minimum size=1.2cm, align=center] (Ann) at (0,0) {Ann};
    \node[draw, circle, minimum size=1.2cm, align=center] (Bob) at (4,0) {Bob};
    \node[draw, circle, minimum size=1.2cm, align=center] (Carol) at (8,0) {Carol};
    \draw[-latex, thick] (Ann) -- (Bob) node[midway, above] {$\alpha$};
    \draw[-latex, thick] (Bob) -- (Carol) node[midway, above] {$\beta$};
    \draw[-latex, thick] (Carol) -- (Bob) node[midway, above] {$\beta$};
\end{tikzpicture}
 \caption{Interaction network among Ann, Bob, and Carol.} \label{figureABC}
\end{figure}
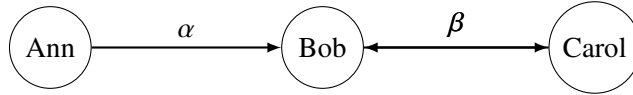

In this example, Ann may suffer from public information.
First, Bob's centrality is $c_B(\ga, G)=1/(1-\ga \beta)$, while Ann's centrality is $c_A(\ga, G)=(1+\ga (\alpha-\beta))/(1-\ga \beta)$. Plugging these into  (\ref{inequalityindividual}) and multiplying it by $(1-\ga\beta)^2$ yields
\begin{equation}
f(\alpha,\beta,\gamma)\equiv (1+\ga (\alpha-\beta))^2-\alpha(2\ga\beta-1)<0.\label{equationABC}
\end{equation}
Let $\mathcal{G}(\ga)\equiv \{(\alpha, \beta) \in [0, 1)^2: f(\alpha,\beta,\gamma)<0\}$ be the set of interaction parameters  $(\alpha, \beta)$ under which Ann suffers from public information, given $\ga$.
It can be shown that $\mathcal{G}(\ga)$ is nonempty if and only if $\ga> (1+\sqrt{5})/4\approx 0.81$, and  $\mathcal{G}(\ga)\subseteq \mathcal{G}(\ga')$ for $\ga 
\leq  \ga'$ (see Figure \ref{figureindividual}).
In Figure \ref{figureindividual}, Ann suffers from public information when $\beta$ is sufficiently large and $\alpha$ is relatively small but not too small. 
\end{example}

The intuition for Proposition \ref{propindividual} is illustrated using Example \ref{exampleABC}.
One might think that Proposition \ref{propindividual} is counterintuitive because Ann wants to align her action with $\theta$ and Bob's action, and the public signal is informative about them.
However, the public signal also informs Bob about what Carol knows (and vice versa), and Ann may suffer from Bob's and Carol's increased knowledge about what their respective neighbor knows.
To see this, compare the current information structure $\mathcal{I}$ with an alternative information structure $\mathcal{I}'$. Under $\mathcal{I}'$, Ann and Bob observe the same signals as $\mathcal{I}$, while Carol observes another private signal  $z\equiv \theta + \varepsilon_z$ instead of $y$.
The noise term $\varepsilon_z$ follows a normal distribution with $\mE[\varepsilon_z]=0$ and $\mathrm{Var}[\varepsilon_z]=\tau_y^{-1}$ and is independent of $\theta$ and all other noise terms.
Thus, relative to $\mathcal{I}$,  agents remain equally informed about $\theta$, Ann and Bob still observe the common signal $y$, but the commonality of information between Bob and Carol is diminished as Carol no longer observes $y$.
The unique linear equilibrium under $\mathcal{I}'$ is
\begin{align}
	a'_{A}&= (1-\ga)(1+\ga \alpha)y+ \ga(1-\alpha + \ga\alpha) x_A;\\
	a'_{B}&=(1-\ga)y+\ga x_B;\\
	a'_{C}&=(1-\ga)z+\ga x_C.
\end{align}
Comparing Ann's equilibrium payoff under $\mathcal{I}$ and $\mathcal{I}'$ yields that Ann is strictly better off under 
$\mathcal{I}'$ if and only if
\begin{equation}
f(\alpha,\beta,\gamma)<(1-\ga \beta)^2\left((1+\ga \alpha)^2+\alpha \right).\label{equationABC2}
\end{equation}
Because the right-hand side of (\ref{equationABC2}) is strictly positive, it follows that Ann suffers from public information \textit{only when} she suffers from Bob and Carol knowing what their neighbor knows.

The detrimental effect on Ann, stemming from the increased commonality of a signal between Bob and Carol, is reminiscent of a mechanism highlighted by \cite{Morris2002-pp}.
Morris and Shin show that when agents' common coordination motive (analogous to $\beta$  in my model) is strong enough and the relative precision of private information is sufficiently high, public information provision increases the action-state distance, i.e., $\mE[(a^*_i-\theta)^2]$.
Similarly, the increased commonality of information between Bob and Carol can make Bob's equilibrium action distant from $\theta$.\footnote{Intuitively, this occurs because Bob and Carol's coordination motive is so large that they choose to be much more responsive to the public signal than they would be when their objectives are to align their actions only with $\theta$.}
This is not desirable for Ann, whose objective is to simultaneously align her action with $\theta$ and Bob's action. 

In sum, public information becomes detrimental to Ann when the loss from Bob's action being more distant from $\theta$ outweighs the benefit from having more precise information about $\theta$ and Bob's action. 
The former loss is more prominent as Bob's coordination motive with Carol increases.
Moreover, Ann's coordination motive with Bob should be relatively small but not too small for the detrimental effect; if it is too small (resp.\ too large), the benefit from knowing more about $\theta$ (resp.\ about Bob's action) becomes dominant.

\subsection{Aggregate Welfare}
Let $W(\tx, \ty, G)\equiv\sum_{i \in N}\mE [u_i(a^*, \theta)]$ denote aggregate welfare in the unique linear equilibrium, and let $\Delta W(\tx, \ty, G)\equiv W(\tx, \ty, G)-W(\tx, 0, G)$ be the welfare effect of public information provision.\footnote{$W(\tx, 0, G) \equiv \sum_{i\in N}U_i(\tx, 0, G)$.}
The result below also follows from (\ref{equationpayoff}).
\begin{prop}\label{propwelfare}
$\Delta W(\tx, \ty, G)<0$ if and only if
 \begin{equation}
 	\sum_{i\in N}c_i(\ga, G)\left((1-d^{in}_i)c_i(\ga, G)+2d^{in}_i\right)<0, \label{inequalitywelfare}
 \end{equation}
 where $d^{in}_i\equiv \sum_{j\neq i} g_{ji}$ is agent $i$'s \textit{in-degree}.
 \end{prop}  
 
For inequality (\ref{inequalitywelfare}) to hold, at least one agent $i$ must have an in-degree $d^{in}_i > 1$ and a Katz-Bonacich centrality $c_i(\gamma, G) > 2d^{in}_i / (d^{in}_i-1)$. This condition requires stronger payoff heterogeneity, as neither regular nor undirected interaction networks satisfy it.\footnote{If $G$ is undirected (i.e., $g_{ij}=g_{ji}$ for all $i, j \in N$), $d^{in}_i = d^{out}_i<1$ for all $i \in N$ by assumption.}
The following example illustrates interaction networks that exhibit such adverse welfare effects.

\begin{example}\label{exampleWelfare}
Consider a directed core-periphery network: $n$ agents are divided into $l \ge 2$ core agents and $m \ge 1$ peripheral agents, so $l+m=n$. 
Both core and peripheral agents intend to coordinate only with core agents; that is, $g_{ij}>0$ only if $j$ is a core agent.
For simplicity, assume $d_i^{out}=\alpha$ if $i$ is peripheral and $d_i^{out}=\beta$ if $i$ is core.
Example  \ref{exampleABC} is a special case of these core-periphery interaction networks with $l=2$ and $m=1$.
Applying (\ref{inequalitywelfare}) to a core-periphery  interaction network yields
 \begin{equation}
 f(\alpha, \beta, \gamma)<-\frac{l}{m}(1-\beta(2\ga\beta-1)). \label{exampleinequalitywelfare}
 \end{equation}
 Let $\mathcal{H}(\ga, l/m)\equiv \{(\alpha, \beta) \in [0, 1)^2: f(\alpha, \beta, \gamma)<-l(1-\beta(2\ga\beta-1))/m\}$ be the set of interaction parameters $(\alpha, \beta)$ that admit the negative welfare effects of public information, given $\ga$ and $l/m$.
It can be shown that $\mathcal{H}(\ga, l/m)$ expands and converges to $\mathcal{G}(\ga)$ as the number of peripheral agents relatively increases, i.e., $\mathcal{H}(\ga, q) \subseteq \mathcal{H}(\ga, p)$ for $p\leq q$ and $\mathcal{H}(\ga, l/m)\rightarrow \mathcal{G}(\ga)$ as $l/m \rightarrow 0$ (see Figure \ref{figurewelfare}).
\begin{figure}[ht!]
    \centering
    \begin{minipage}{0.3\textwidth}
        \centering
        \includegraphics[width=\linewidth]{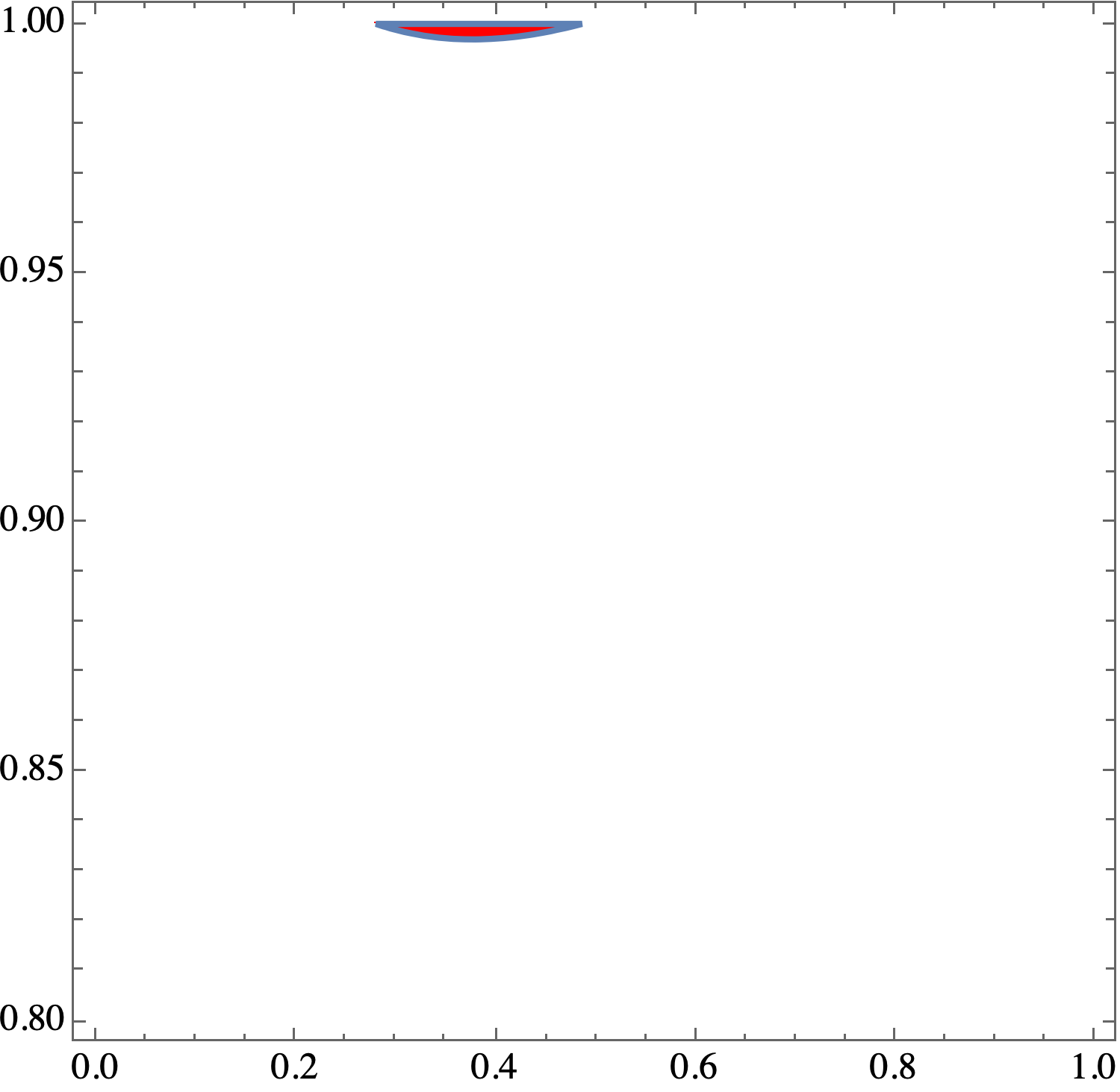}  
        \subcaption{$l/m= 0.5$}
    \end{minipage}%
      \hspace{0.5cm}
    \begin{minipage}{0.3\textwidth}
        \centering
        \includegraphics[width=\linewidth]{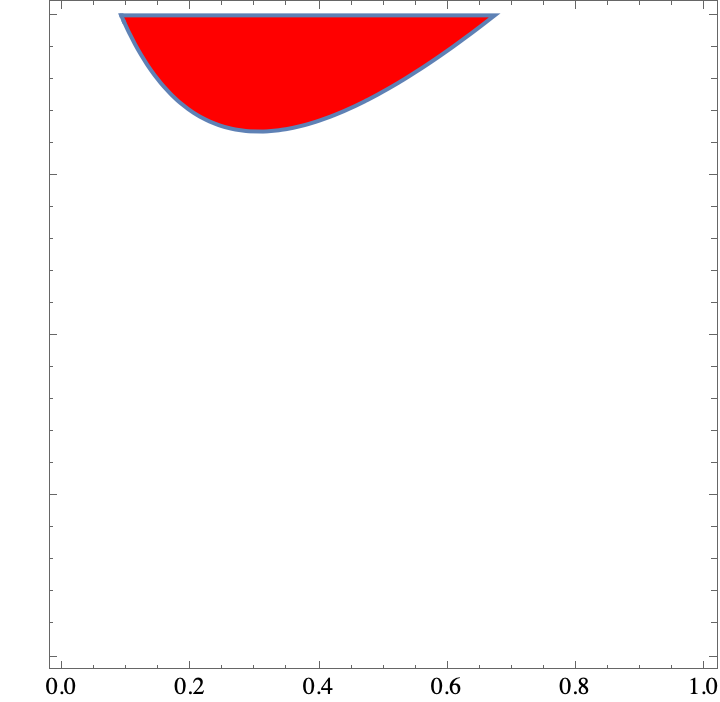}  
        \subcaption{$l/m= 0.2$}
    \end{minipage}%
      \hspace{0.5cm}
    \begin{minipage}{0.3\textwidth}
        \centering
        \includegraphics[width=\linewidth]{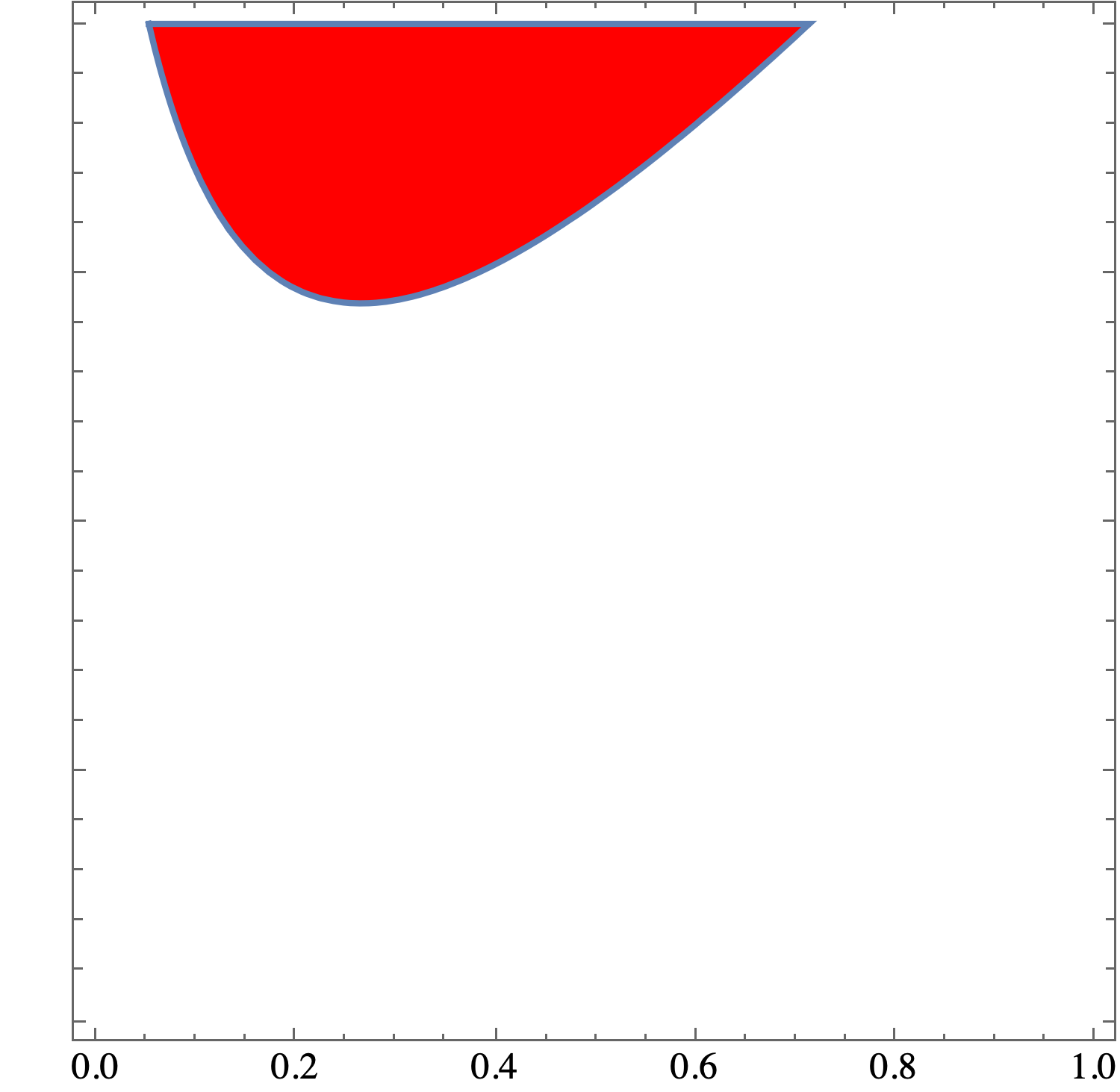}  
        \subcaption{$l/m= 0.1$}
    \end{minipage}
        \caption{ $\mathcal{H}(\ga, l/m)$ for $\ga=0.9$ and $l/m\in \{0.5, 0.2, 0.1\}$.}\label{figurewelfare}
\end{figure}
\end{example}

The basic intuition for Proposition \ref{propwelfare} parallels that for Proposition \ref{propindividual}. One addition is that in Example \ref{exampleWelfare}, the in-degree of core agents reflects the extent to which peripheral agents target them for coordination. Distortion of core agents' actions from the payoff state, resulting from their response to public information, imposes a larger aggregate loss when many peripheral agents attempt to coordinate with these distorted actions. Thus, for the total loss to peripheral agents to outweigh the total gain accrued by the core agents, at least some core agents must have a sufficiently large in-degree.

Proposition \ref{propwelfare} implies a stark non-monotonic relationship between network connectivity and the social value of public information.
To see this, I first contrast this result with the benchmark of regular networks, where all agents share the same out-degree, i.e., $d_i^{out} = d$ for all $i \in N$.
In this environment, denser connectivity unambiguously increases the social value of public information.
The left-hand side of inequality (\ref{inequalitywelfare}) simplifies to
\begin{equation}
n(1+d-2\ga d^2)/(1-\ga d)^2,
\end{equation}
which is positive and strictly increasing in $d$.\footnote{Suppose $d_i^{out}=d$ for all $i\in N$.
Then, the term $c_i^2(\ga, G)-\sum_{j \neq i}g_{ij}c_j(\ga, G)\left(c_j(\ga, G)-2\right)$ simplifies to $(1+d-2\ga d^2)/(1-\ga d)^2$.
Differentiating this ratio with respect to $d$ yields $((1-\ga d)+2\ga(1-d))/(1-\ga d)^3$, which is strictly positive for all $d\in [0,1)$.}
Thus, under the uniform interaction, the welfare effect of public information monotonically increases with network connectivity.
This confirms the standard intuition that as agents become more incentivized to coordinate, the public signal becomes increasingly valuable.

However, this intuition breaks down under heterogeneous interactions, and increasing connectivity can turn a beneficial public signal into a detrimental one.
The following corollary formalizes this statement.
\begin{cor}
\label{propmonotonicity}
	Suppose $n \geq 3$. Then, there exist $\gamma \in (0, 1)$ and two interaction networks $G$ and $G'$ with $G\leq G'$ such that
    \begin{equation}
    \Delta W(\tau_x, \tau_y, G') < 0 < \Delta W(\tau_x, \tau_y, G).\footnote{$G\leq G'$ if $g_{ij}\leq g'_{ij}$ for all $i, j \in N$.}
    \end{equation}
\end{cor}

The proof constructs the core-periphery network $G'$ by adding weights to an empty network $G$ to concentrate connectivity on a small subset of agents, rather than densifying the network uniformly. This selective increase generates a dense core with disproportionately high centrality. By linking the other agents to this core with small but strictly positive weights, the construction exposes them to the core's excessive volatility, while the direct benefits of coordination remain limited. Consequently, when public information is imprecise relative to private information (i.e., $\gamma$ is sufficiently high), this construction leads to a net reduction in aggregate welfare.

\subsection{Endogenous Information Sharing}
The obtained results have implications for the efficiency of endogenous information sharing.
Consider an extended model in which agent 1, initially possessing a signal $y$, chooses whether to make $y$ public or to keep it private before signal realizations.
Agent $1$'s choice selects the information structure of the beauty contest game: If $y$ is made public, the information structure will be $\mathcal{I}$ as in the main analysis; if $y$ is kept private, the information structure will be $\mathcal{I}^{\dagger}$ under which agent 1 observes two private signals, $x_1$ and $y$, while each other agent $i \neq 1$ observes only her private signal $x_i$.\footnote{It is implicitly assumed that other signals, such as $x_1$, are not shareable. One might justify this assumption by considering $x_1$ as tacit knowledge (e.g., personal experience) and $y$ as explicit knowledge (e.g., an analyst report).}
There is no cost to share information.

The two-stage game is solved by backward induction. The analysis requires comparing equilibrium payoffs under public sharing with those under private retention of the signal.
As derived in Section \ref{sectionproofeqmshare}, the equilibrium actions in the subgame following a choice to keep the signal private are
\begin{align}
	a^\dagger_{1}=  (1-\ga)y + \ga x_1, \quad \text{and}\quad a^\dagger_{i}= x_i\ \text{for each } i \neq 1.
\end{align}
Agent $1$'s equilibrium payoff and aggregate welfare are then computed as
\begin{align}
	U^\dagger_1(\tx,\ty, G)&=\tx^{-1}(1-\ga)+U_1(\tx, 0, G);\\
	W^\dagger(\tx,\ty, G)&=\tx^{-1}(1-\ga)(1+d_1^{in})+W(\tx, 0, G).
\end{align}
Combining these expressions with Proposition \ref{propindividual} and \ref{propwelfare} yields the following result.

\begin{figure}[ht!]
    \centering
    \begin{minipage}{0.3\textwidth}
        \centering
        \includegraphics[width=\linewidth]{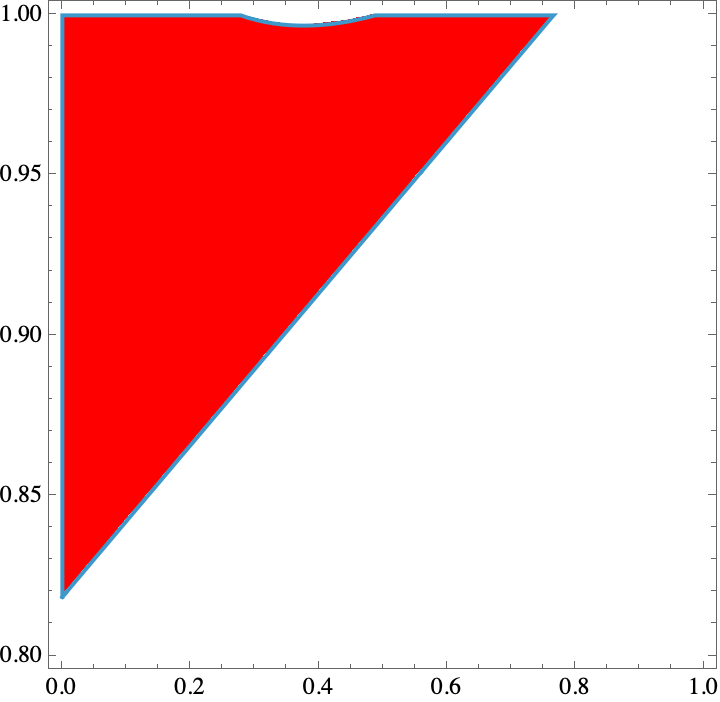}  
        \subcaption{$m=44$}
    \end{minipage}%
      \hspace{0.5cm}
    \begin{minipage}{0.3\textwidth}
        \centering
        \includegraphics[width=\linewidth]{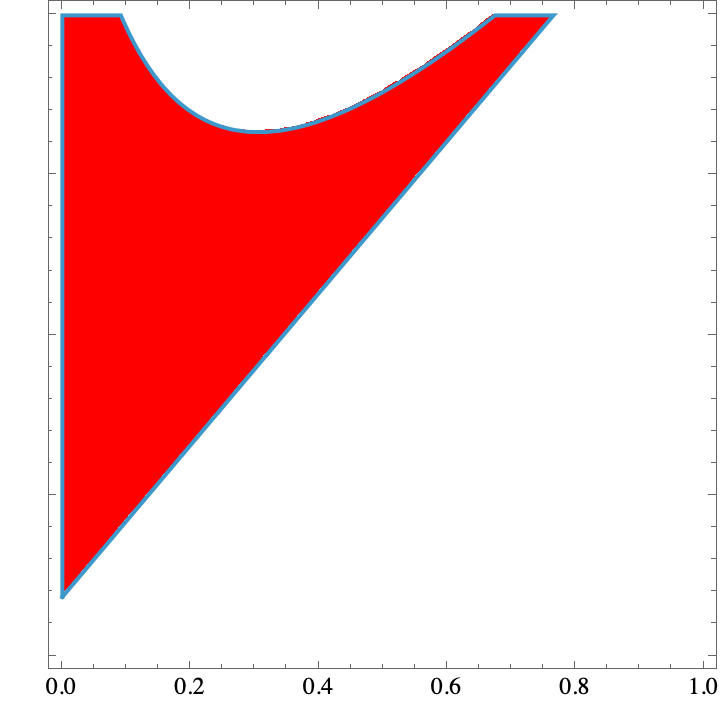}  
        \subcaption{$m=55$}
    \end{minipage}%
      \hspace{0.5cm}
    \begin{minipage}{0.3\textwidth}
        \centering
        \includegraphics[width=\linewidth]{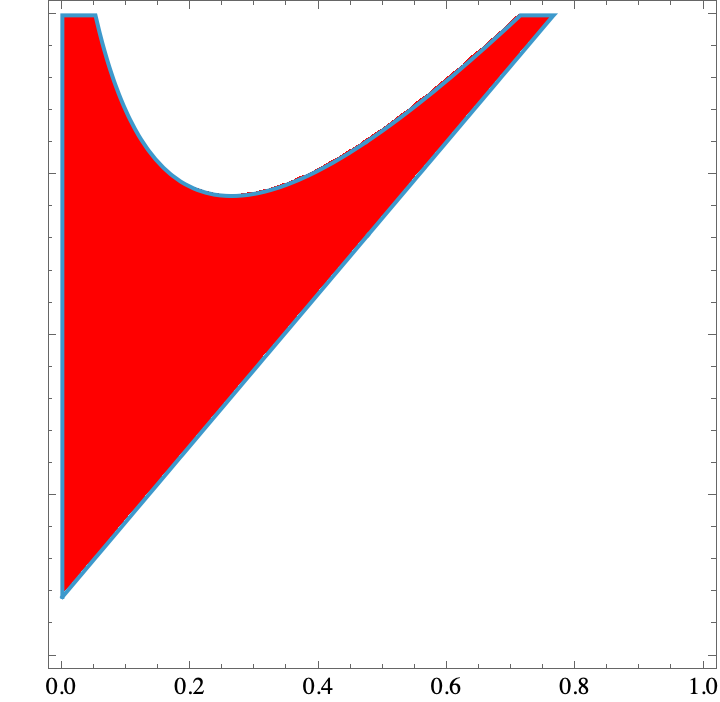}  
        \subcaption{$m=60$}
    \end{minipage}
        \caption{$\mathcal{J}(\ga, l, m)$ for $\ga=0.9$, $n=66$, and $m\in \{44, 55, 60\}$.}\label{figureinfoshare}
\end{figure}

\begin{prop}\label{propinfoshare}
The equilibrium degree of information sharing is inefficiently low, i.e., 
$U_1(\tx,\ty, G)<U^\dagger_1(\tx,\ty, G)$ and $W^\dagger(\tx,\ty, G)<W(\tx,\ty, G)$, if and only if the following two conditions hold:
\begin{gather}
    c_1^2(\ga, G)-\sum_{j \neq 1}g_{1j}c_j(\ga, G)\left(c_j(\ga, G)-2\right) < 1;  \label{inequalityinfoshare1}\\
    1+d^{in}_{1} < \sum_{i\in N}c_i(\ga, G)\left((1-d_i^{in})c_i(\ga, G)+2d_i^{in}\right).
 \label{inequalityinfoshare2}
\end{gather}
\end{prop}
In words, the equilibrium degree of information sharing becomes inefficiently low if and only if: (i) agent 1 is harmed by making her information public, which typically occurs if her Katz-Bonacich centrality is sufficiently small relative to an aggregate measure of her neighbors’ centralities, and (ii) aggregate welfare would nonetheless increase if the information were shared, which typically occurs if the interaction network does not have agents with very high in-degrees and centralities.
For instance, consider the core-periphery interaction networks introduced in Example \ref{exampleWelfare}.
Assuming that agent 1 is peripheral, inequalities (\ref{inequalityinfoshare1}) and (\ref{inequalityinfoshare2}) simplify to
 \begin{equation}
  f(\alpha, \beta, \gamma)< (1-\beta\gamma)^2<m f(\alpha, \beta, \gamma)+l(1-\beta(2\ga\beta-1)). \label{exampleinequalityinfoshare}
 \end{equation}
 Let $\mathcal{J}(\ga, l, m)\equiv \{(\alpha, \beta) \in [0, 1)^2: f(\alpha, \beta, \gamma)< (1-\beta\gamma)^2<m  f(\alpha, \beta, \gamma)+l(1-\beta(2\ga\beta-1))\}$ be the set of interaction parameters $(\alpha, \beta)$ that admit inefficiently low equilibrium information sharing, given $\ga$, $l$,  and $m$.
 Because agent 1's preference over information structures is ``more aligned'' with peripheral agents than with core agents, $\mathcal{J}(\ga, l,m)$ shrinks as the number of peripheral agents relatively increases, i.e., $\mathcal{J}(\ga, n-m', m') \subseteq \mathcal{J}(\ga, n-m,m)$ for $m \leq m'$ (see Figure \ref{figureinfoshare}).\footnote{The formal argument is as follows.  Fix $\ga$ and take any $(\alpha, \beta)$ that satisfies $f(\alpha, \beta, \gamma)< (1-\beta\gamma)^2$.
The right-hand side of (\ref{exampleinequalityinfoshare}) is linear in $m$ with slope $f(\alpha, \beta, \gamma)-(1-\beta(2\ga\beta-1))$ by substituting $l=n-m$.  
  This slope is bounded from above by
\begin{align*}
f(\alpha, \beta, \gamma)-1+2\ga\beta^2-\beta
&\leq f(\alpha, \beta, \gamma)-1+ 2\ga\beta-\beta \\
&\leq f(\alpha, \beta, \gamma)-1 + 2\ga\beta-\ga \beta\\
 &=f(\alpha, \beta, \gamma)-1+ \ga\beta\\
  &< 0.
 \end{align*}   
The last strict inequality follows from the initial assumption. Thus, the right-hand side of (\ref{exampleinequalityinfoshare}) strictly decreases with $m$ whenever $f(\alpha, \beta, \gamma)< (1-\beta\gamma)^2$, which proves the claim.}

Intuitively, an agent may withhold her information to prevent her neighbors from excessively responding to her signal due to their strong coordination motives. Of course, keeping information private is not necessarily socially desirable, especially when other agents would benefit from the commonality of information. Thus, the resulting inefficiency can be substantial in the core-periphery interaction networks when the core is relatively large. This finding presents an interesting contrast to the case of public information provision, where the adverse welfare effects can be pronounced when the periphery is relatively large.

\section{Discussion}
\subsection{Marginal Value of Public Information}
While the preceding analysis identifies conditions for discrete welfare losses, a more calibrated policy requires characterizing the marginal value of transparency. This section examines the sensitivity of aggregate welfare to a local increase in the public signal precision $\tau_y$.

Recall that, based on the equilibrium characterization, aggregate welfare is simplified to
\begin{equation}
W(\tau_x, \tau_y, G) =  \tau_x^{-1}(1-\gamma) S(\gamma)+C,
\end{equation}
where $C$ is a constant independent of $\tau_y$, and $S(\gamma)$ captures the network-dependent welfare component:
\begin{equation}
S(\gamma) \equiv \sum_{i \in N} c_i(\gamma, G) \left( (1-d_i^{in})c_i(\gamma, G) + 2d_i^{in} \right).
\end{equation}

Increasing transparency creates a tension between signal reliance and network feedback. On one hand, a higher $\tau_y$ increases the baseline weight $1-\gamma$ placed on the public signal. This scaling effect amplifies the distortion captured by $S(\gamma)$. On the other hand, it reduces the discount factor $\gamma$ governing centrality. This dampening effect shrinks the centrality vector, attenuating the magnitude of the distortion.
The \textit{centrality sensitivity vector} $\mathbf{c} ^\mathbf{c}(\gamma, G) \equiv (I - \gamma G)^{-1} \mathbf{c}(\gamma, G)$ captures the sensitivity of centrality to variations in $\gamma$, as the derivative of centrality vector with respect to $\gamma$ is proportional to the difference $\mathbf{c}^\mathbf{c}(\gamma, G) - \mathbf{c}(\gamma, G)$.\footnote{Differentiating $\mathbf{c} = (I - \gamma G)^{-1}\mathbf{1}$ yields $\partial \mathbf{c}/\partial \gamma = (I - \gamma G)^{-1} G \mathbf{c}$. Using the identity $\mathbf{c} = \mathbf{1} + \gamma G \mathbf{c}$, I substitute $G \mathbf{c} = \gamma^{-1}(\mathbf{c} - \mathbf{1})$ to obtain $\partial \mathbf{c}/\partial \gamma = \gamma^{-1} \left( (I - \gamma G)^{-1}\mathbf{c} - (I - \gamma G)^{-1}\mathbf{1} \right)$. By definition, this simplifies to $\gamma^{-1}(\mathbf{c}^\mathbf{c} - \mathbf{c})$.}

\begin{prop}\label{prop:marginal_welfare}
The marginal value of public information is negative, i.e., $\partial W/\partial \tau_y < 0$, if and only if
\begin{equation}\label{eq:marginal_condition}
S(\gamma) < (1-\gamma) S'(\gamma),
\end{equation}
where 
\begin{equation}
S'(\gamma) = 2 \gamma^{-1}\sum_{i \in N} \left( c^\mathbf{c}_i(\gamma, G) - c_i(\gamma, G) \right) \left( (1-d_i^{in})c_i(\gamma, G) + d_i^{in} \right).
\end{equation}
\end{prop}

Inequality (\ref{eq:marginal_condition}) imposes a more stringent condition than the one for a discrete welfare loss identified in Proposition \ref{propwelfare}. While a discrete loss arises whenever the aggregate distortion is negative, a marginal welfare reduction requires this distortion to be severe enough to overwhelm the dampening effect.\footnote{Note that when the network exhibits a discrete welfare loss (i.e., $S(\gamma) < 0$), the sensitivity term $S'(\gamma)$ is typically negative. This occurs because agents who have high centrality and high in-degree are also the most sensitive to changes in $\gamma$, implying a large gap between their centrality and its sensitivity (i.e., $c_i^\mathbf{c}(\gamma, G) - c_i(\gamma, G)$).} It is therefore possible for a network to exhibit a discrete welfare loss yet benefit from marginal increases in transparency if the dampening effect is sufficiently strong. This distinction vanishes in the limit where public information becomes negligible (i.e., $\gamma \to 1$). In this case, the sensitivity term on the right-hand side of (\ref{eq:marginal_condition}) converges to zero, reducing the condition to $S(1) < 0$. This confirms that an interaction network generating a discrete welfare loss must also make the initial introduction of public information marginally detrimental. 

\subsection{Alternative Payoff Specification}
This section generalizes the payoff function to disentangle the intrinsic motive for coordination from the structural constraints of the interaction network. Agent $i$'s payoff function is given by
\begin{equation}\label{alternative_payoff}
u_{i}(a, \theta)= -\left(1-r_i\right)(a_{i}-\theta)^{2} - r_i\sum_{j \neq i}g_{ij}(a_{i}-a_{j})^{2},
\end{equation}
The parameter $r_i \in (0, 1)$ captures the relative intensity of coordination motives against adaptation motives.
The corresponding first-order condition yields
\begin{equation}\label{FOC_alt}
a_i= \left(1-\tilde{d}_i^{out}\right)\mathbb{E}_i[\theta]+ \sum_{j \neq i}\tilde{g}_{ij}\mathbb{E}_i[a_{j}],
\end{equation}
where $\tilde{g}_{ij}$ denotes the $(i,j)$-th element of the \textit{normalized interaction network} $\tilde{G}$, defined by
\begin{equation}
\tilde{g}_{ij} \equiv \frac{r_i g_{ij}}{1-r_i + r_i d_i^{out}},
\end{equation}
and $\tilde{d}_i^{out} \equiv \sum_{j \neq i} \tilde{g}_{ij}$ is the normalized out-degree of agent $i$.

The normalized interaction weight $\tilde{g}_{ij}$ is strictly increasing in both the pairwise link strength $g_{ij}$ and the coordination intensity $r_i$, but strictly decreasing in the out-degree $d_i^{out}$.\footnote{This captures a ``dilution effect'': as an agent's neighborhood expands, the relative influence of any single peer $j$ diminishes, effectively spreading the agent's coordination capacity across a wider set of interactions.} Because equation (\ref{FOC_alt}) takes the same form as the best-response condition (\ref{FOC}) in the baseline model with $\tilde{G}$ replacing $G$, and $\tilde{G}$ is row-substochastic, the game admits a unique linear equilibrium characterized by the Katz-Bonacich centrality on $\tilde{G}$. This equivalence allows the welfare analysis of the baseline model to be naturally extended to this generalized setting.


\begin{prop}\label{propwelfare_alt}
The welfare effect of public information provision is negative, i.e., $\Delta W(\tau_x, \tau_y, G) < 0$, if and only if
\begin{equation}
\sum_{i \in N} c_i(\gamma, \tilde{G}) \left( \left(K_i - d_i^{in, r}\right)c_i(\gamma, \tilde{G}) + 2d_i^{in, r} \right) < 0, \label{inequalitywelfare_alt}
\end{equation}
where $K_i \equiv 1-r_i+r_i d_i^{out}$ and $d_i^{in, r} \equiv \sum_{j \neq i} r_j g_{ji}.$
\end{prop}
In inequality (\ref{inequalitywelfare_alt}), the term $K_i$ represents agent $i$'s total payoff weight, capturing the combined magnitude of losses from misaligned adaptation and coordination. The term $d_i^{in, r}$ measures agent $i$'s effective influence, derived by weighting the coordination motives directed at $i$ by the intensities of her neighbors. Thus, aggregate welfare decreases when disproportionately central agents exert high effective influence relative to their total payoff weight.

This result implies that neither heterogeneity in directionality (i.e., $r_i =r$ and $d_i^{out}= 1$, but $g_{ij}$ varies) nor heterogeneity in intensity (i.e., $g_{ij}$ is uniform, but $r_i$ varies) is sufficient to generate negative welfare effects in isolation. Instead, the detrimental effect requires the interplay of these two factors. The following corollary formalizes this necessity.\begin{cor}\label{prop:dual_heterogeneity}The welfare effect of public information provision is positive if either (1) $r_i=r$ and $d_i^{out}=1$ for all $i \in N$, or (2) $g_{ij}=1/(n-1)$ for all $i \neq j$.\end{cor}

\subsection{Efficient Use of Information}
To analyze how the welfare losses identified above relate to agents' use of information, I characterize the \textit{efficient use of information} following the approach of \cite{Angeletos2007-uq}. This efficient benchmark is defined as the strategy profile that maximizes aggregate welfare, subject to the constraint that agents must condition their actions solely on their own decentralized information sets. By keeping the informational friction constant while internalizing payoff interdependencies, this comparison isolates the specific inefficiency driven by the gap between private and social incentives.

In my framework, the efficient strategies are equivalent to the equilibria of a beauty contest game played on a modified network. Specifically, the following first-order condition characterizes the efficient strategies:
\begin{equation}\label{efficient_FOC}
a_i = \left(1 - d_i^{F}\right) \mathbb{E}_i[\theta] + \sum_{j \neq i} g_{ij}^F \mathbb{E}_i[a_j],
\end{equation}
where $d_i^{F} \equiv \sum_{j \neq i} g_{ij}^F$ and $G^F=[g_{ij}^F]$ is the \textit{fictitious interaction network} defined by
\begin{equation}
g_{ij}^F \equiv \frac{g_{ij} + g_{ji}}{1 + d_i^{in}}.
\end{equation}
This fictitious network captures the wedge between private and social coordination motives. 
The numerator strengthens bilateral coordination incentives by internalizing both sides' payoffs, while the denominator scales these bilateral motives down more strongly for high in-degree agents than for low in-degree ones.
 Using this equivalence, Proposition \ref{prop:efficient_use} presents a unique linear efficient strategy.

\begin{prop}\label{prop:efficient_use}
There exists a unique linear efficient strategy profile in which agent $i$'s strategy is
\begin{equation}
a_i^{F} = (1-\gamma)c_i(\gamma, G^F) y + (1 - (1-\gamma)c_i(\gamma, G^F)) x_i,
\end{equation}
where $c_i(\gamma, G^F)$ is $i$'s Katz-Bonacich centrality in $G^F$.
\end{prop}

This result contrasts with the findings of \cite{Angeletos2007-uq}. In their analysis of symmetric settings, they show that when the social degree of coordination exceeds the private one, agents \textit{under-respond} to public information in equilibrium.\footnote{In \cite{Angeletos2007-uq}, the private degree of coordination represents the weight assigned to the average action in the equilibrium first-order condition. In contrast, the social degree of coordination represents the corresponding weight in the planner's first-order condition.} In contrast, in the present framework, the direction of inefficiency is governed by network topology; in particular, agents \textit{over-respond} to public information in equilibrium whenever their private centrality exceeds their social centrality, $c_i(\gamma, G) > c_i(\gamma, G^F)$.

\begin{figure}[ht]\centering
\begin{tikzpicture}[scale=0.8, every node/.style={scale=0.9}]
    \node[draw, circle, minimum size=1.0cm, align=center] (Ann) at (0,0) {Ann};
    \node[draw, circle, minimum size=1.0cm, align=center] (Bob) at (5,0) {Bob};
    \node[draw, circle, minimum size=1.0cm, align=center] (Carol) at (10,0) {Carol};

    \draw[-latex, thick] (Ann) to[bend left=30] node[midway, above] {$\alpha$} (Bob);

    \draw[-latex, thick] (Bob) to[bend left=30] node[midway, below] {$\frac{\alpha}{1 + \alpha + \beta}$} (Ann);

    \draw[-latex, thick] (Bob) to[bend left=30] node[midway, above] {$\frac{2\beta}{1 + \alpha + \beta}$} (Carol);

    \draw[-latex, thick] (Carol) to[bend left=30] node[midway, below] {$\frac{2\beta}{1 + \beta}$} (Bob);
\end{tikzpicture}
\caption{The fictitious interaction network corresponding to Figure \ref{figureABC}.} \label{figureABC_fictitious}
\end{figure}
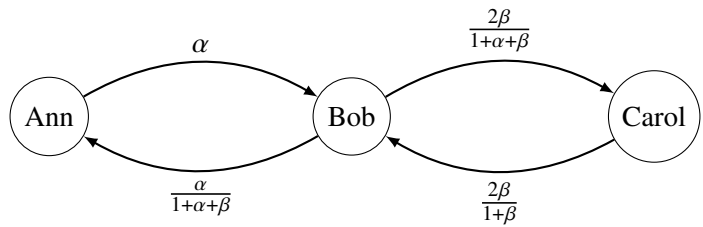

To illustrate the locus of inefficiency, consider Example \ref{exampleABC} of Ann, Bob, and Carol. Figure \ref{figureABC_fictitious} presents the corresponding fictitious network. In this fictitious network, Bob's weight on Ann increases with $\alpha$, reflecting the planner's internalization of the unilateral externality Bob exerts on Ann. Conversely, Bob's weight on Carol decreases with $\alpha$, thereby weakening the coordination loop between Bob and Carol as Ann's dependence on Bob intensifies.

This structural transformation identifies Bob as the primary source of inefficiency.
Consider a numerical specification with a strong core ($\beta = 0.95$), a weakly dependent periphery ($\alpha = 0.2$), and high relative precision of private signals ($\gamma=0.95$).
Under these parameters, Bob's centrality strictly decreases from a private level of $c_B(\gamma, G) \approx 10.26$ to a social level of $c_B(\gamma, G^F) \approx 9.35$.\footnote{In the fictitious network, the weights incident to Bob are $g_{BA}^F = \alpha/(1+\alpha+\beta)$ and $g_{BC}^F = 2\beta/(1+\alpha+\beta)$, while the return links are $g_{AB}^F=\alpha$ from Ann and $g_{CB}^F = 2\beta/(1+\beta)$ from Carol. Bob's centrality satisfies the recursive condition $c_B = 1 + \gamma g_{BA}^F c_A + \gamma g_{BC}^F c_C$. Substituting the expressions for neighbors' centralities, $c_A = 1 + \gamma \alpha c_B$ and $c_C = 1 + \gamma g_{CB}^F c_B$, into Bob's equation yields
$$c_B = 1 + \gamma g_{BA}^F (1 + \gamma \alpha c_B) + \gamma g_{BC}^F (1 + \gamma g_{CB}^F c_B).$$
Grouping the terms with $c_B$ on the left-hand side and solving yields the closed-form solution:
$$c_B(\gamma, G^F) = \frac{1 + \gamma (g_{BA}^F + g_{BC}^F)}{1 - \gamma^2 (\alpha g_{BA}^F + g_{BC}^F g_{CB}^F)}.$$
For parameters $\beta=0.95, \alpha=0.2$, and $\gamma=0.95$, $c_B(\gamma, G^F) =808275/86408$, which is strictly lower than the private centrality $c_B(\gamma, G)=400/39$.}
Thus, Bob over-responds to public information relative to the efficient benchmark. This localized excess coordination creates a negative externality for Ann, mirroring the classic tension between private and social incentives highlighted by \cite{Morris2002-pp}.

\section{Conclusion}
This paper investigates how the structure of interaction networks determines the welfare effects of public information. I show that the sign of the welfare effect depends on the interplay between Katz-Bonacich centrality and aggregate influence (i.e., in-degree). In the tiered core-periphery architectures typical of OTC markets, a small, dense core is sufficiently central and influential that its excessive responsiveness to public signals generates a volatility externality, outweighing the coordination benefits for the periphery.
This mechanism also uncovers a strategic friction in information sharing that leads to the under-provision of public information: agents may rationally withhold signals to avoid exposure to the volatility induced by their highly connected peers.

These findings imply that the value of transparency is non-monotonic in network connectivity. While uniform connectivity enhances the value of public information, the emergence of dominant hubs can reverse it. Future research could investigate how these forces shape endogenous network formation, particularly whether agents strategically limit connections to mitigate the volatility externality identified here.

\appendix
\section*{Appendix: Proofs} 
\addcontentsline{toc}{section}{Appendix: Proofs} 
\setcounter{section}{1} 
\label{sectionproof}
\subsection{Proposition \ref{propequilibrium}}\label{sectionproof1}
The remaining part of the proof is to show that the sum of the equilibrium slopes must be one in any linear equilibrium.
Suppose $a_i=b_i^yy+b_i^xx_i$ for all $i \in N$.
Substituting these linear strategies into (\ref{FOC}) yields
\begin{align*}
b^y_i y + b^x_i x_i&=(1-d_i^{out})\left((1-\ga)y+  \ga x_i\right)+ \sum_{j \neq i}g_{ij}\left(b_j^y y + b_j^x((1-\ga)y+\ga x_i)\right)\\
&= \left((1-d_i^{out})(1-\ga) + \sum_{j \neq i}g_{ij}b^y_j+(1-\ga)\sum_{j \neq i}g_{ij}b^x_j\right)y+\ga \left(1-d_i^{out} + \sum_{j \neq i}g_{ij}b^x_j\right)x_i,
\end{align*}
where the first equality follows from $\mE_i[\theta]=(1-\ga)y+  \ga x_i$ and $\mE_i[a_j]=b_j^y y+  b_j^x \left((1-\ga)y+  \ga x_i\right)$.
By matching the coefficients, the equilibrium slopes must satisfy
\begin{align*}
b^y_i - \sum_{j \neq i}g_{ij}b^y_j-(1-\ga)\sum_{j \neq i}g_{ij}b^x_j&= (1-\ga)(1-d_i^{out});  \\
b^x_i -\ga \sum_{j \neq i}g_{ij}b^x_j&= \ga (1-d_i^{out}) .
\end{align*}
Summing these two equations yields
\begin{equation*}
b^y_i+b^x_i -\sum_{j \neq i}g_{ij}\left(b^y_j+b^x_j\right)= 1-d_i^{out}.
\end{equation*}
In matrix notation, \begin{equation*}
(I - G)(\mb^y+\mb^x) =(I - G)\mone,\end{equation*}
where $\mb^y=(b^y_1, \ldots, b^y_n)'$ and $\mb^x=(b^x_1, \ldots, b^x_n)'$ are $n$-dimensional column vectors.
Because the inverse matrix $(I-G)^{-1}$ exists by assumption, $\mb^y+\mb^x=(I-G)^{-1}(I-G)\mone=\mone$.
Thus, any linear equilibrium must satisfy $b_i^y+b_i^x=1$ for all $i\in N$, as desired.
\qed

\subsection{Proposition \ref{propindividual}}
By Proposition \ref{propequilibrium}, $X_i=(1-(1-\ga)c_i(\ga, G))\varepsilon_i+(1-\ga)c_i(\ga, G)\varepsilon_y$ does not depend on $\theta$, and the ex-ante expected payoff is well defined. I simplify $\mE[X^2_i]$, $\mE[X_iX_j]$, and $\sum_{j\neq i}g_{ij}\mE[X_iX_j]$ as follows:
\begin{align*}
	\mE[X^2_i] &= \mE[((1-b_i)\varepsilon_{i}+b_i \varepsilon_{0})^2]\\
	&= (1-b_i)^2\tx^{-1}+b_i^2\ty^{-1}\\
	&= \tx^{-1}(1-(1-\ga)c_i(\ga, G))^2+\ty^{-1}\left((1-\ga)c_i(\ga, G)\right)^2\\
    &= (\tx^{-1}+\ty^{-1})(1-\ga)^2 c^2_i(\ga, G)-2\tx^{-1}(1-\ga)c_i(\ga, G)+\tx^{-1}\\
	&= \tx^{-1}(1-\ga)c^2_i(\ga, G)-2\tx^{-1}(1-\ga)c_i(\ga, G)+\tx^{-1}\\
	&= \tx^{-1}(1-\ga)c_i(\ga, G) (c_i(\ga, G)-2)+\tx^{-1};\\
	\mE[X_iX_j] &= \mE[((1-b_i)\varepsilon_{i}+b_i \varepsilon_{0})((1-b_j)\varepsilon_{j}+b_j \varepsilon_{0})]\\
	&= b_ib_j\ty^{-1}\\
	&= \ty^{-1}(1-\ga)^2c_i(\ga, G)c_j(\ga, G);\\
	\sum_{j\neq i}g_{ij}\mE[X_iX_j] &= \ty^{-1}(1-\ga)^2c_i(\ga, G)\sum_{j\neq i}g_{ij}c_j(\ga, G)\\
	&= \ty^{-1}(1-\ga)^2c_i(\ga, G) \ga^{-1}(c_i(\ga, G)-1)\\
	&= \tx^{-1}(1-\ga)c_i(\ga, G)(c_i(\ga, G)-1).
\end{align*}
The second last equality follows from the identity $\mc(\ga, G)=\mone+\ga G \mc(\ga, G)$.

Using these expressions, agent $i$'s equilibrium payoff simplifies to
\begin{align*}
	U_i(\tx, \ty, G) &= -\left(1-d_i^{out}\right)\mE[X^2_i] - \sum_{j \neq i}g_{ij}\mE[(X_i-X_j)^2]\\
    &=-\mE[X^2_i]- \sum_{j \neq i}g_{ij}\mE[X_j^2]+2 \sum_{j \neq i}g_{ij}\mE[X_i X_j]\\
    &= \tx^{-1}(1-\ga)\left(-c_i(\ga, G) (c_i(\ga, G)-2)+2c_i(\ga, G)(c_i(\ga, G)-1)\right)-\sum_{j \neq i}g_{ij}\mE[X_j^2]-\tx^{-1}\\
    &=  \tx^{-1}(1-\ga)\left(c_i^2(\ga, G)-\sum_{j \neq i}g_{ij}c_j(\ga, G)\left(c_j(\ga, G)-2\right)\right)-\tx^{-1}(1+d_i^{out}).
\end{align*}
Because $U_i(\tx, 0, G)=-\tx^{-1}(1+d_i^{out})$, it follows that $\Delta U_i(\tx, \ty, G)<0$ if and only if $c_i^2(\ga, G)-\sum_{j \neq i}g_{ij}c_j(\ga, G)\left(c_j(\ga, G)-2\right)<0$, as desired.\qed

\subsection{Claims in Example \ref{exampleABC}}\label{sectionproofexample1}
\begin{claim}
$\Delta U_{A}(\tx,\ty, G)<0$ if and only if  $f(\alpha,\beta,\gamma)<0$.
\end{claim}
\noindent The centralities of Ann and Bob are derived using the self-referentiality:
$c_A(\ga, G)=1+\ga \alpha c_B(\ga, G)$, $c_B(\ga, G)=1+\ga \beta c_C(\ga, G)$, and $c_C(\ga, G)=1+\ga \beta c_B(\ga, G)$.
Solving these simultaneous equations yields $c_A(\ga, G)=(1+\ga (\alpha-\beta))/(1-\ga \beta)$ and $c_B(\ga, G)=c_C(\ga, G)=1/(1-\ga \beta)$.
For Ann, plugging these into the terms inside the large parentheses of (\ref{equationpayoff})  yields
\begin{align*}
  c_A^2(\ga, G)-\alpha c_B(\ga, G)\left(c_B(\ga, G)-2\right) &=  (1-\ga\beta)^{-2}\left((1+\ga (\alpha-\beta))^2-\alpha (1-2(1-\ga\beta))\right)\\
  &=(1-\ga\beta)^{-2}\left((1+\ga (\alpha-\beta))^2-\alpha(2\ga\beta-1)\right)\\
  &=(1-\ga\beta)^{-2}f(\alpha,\beta,\gamma).
\end{align*}
Hence, (\ref{inequalityindividual}) holds for Ann if and only if $f(\alpha,\beta,\gamma)<0$. \qed

\begin{claim}
$\mathcal{G}(\ga)$ is non-empty if and only if $\ga> (1+\sqrt{5})/4$. 
\end{claim}
\noindent
It suffices to show that there exists a triplet $(\alpha,\beta,\gamma)$ such that $f(\alpha,\beta,\gamma)<0$ if and only if $\ga> (1+\sqrt{5})/4$.
Define a quadratic  function $h(\alpha,\ga)\equiv\inf_{\beta\in [0, 1)} f(\alpha, \beta,\gamma)$.
Simple algebra shows
\begin{align*}
	h(\alpha,\ga)	&= f(\alpha, 1,\gamma) \\
	&= (1+\ga (\alpha-1))^2-\alpha(2\ga-1)\\
	&= \ga^2 \alpha^2+2\ga(1-\ga)\alpha +(1-\ga)^2-2\ga\alpha+\alpha\\
	&= \ga^2 \alpha^2+(1-2\ga^2)\alpha +(1-\ga)^2.
\end{align*}
Because $h(0, \ga) > 0$ and $h(1, \ga) > 0$ for any $\ga \in (0,1)$, the quadratic function $h(\alpha,\ga)$ can be strictly negative for some $\alpha \in [0,1)$ if and only if its minimum is negative. This requires two conditions to hold jointly: (i) the vertex must lie in the interval $(0,1)$, i.e., $\ga > 1/\sqrt{2}$, and (ii) the value at the vertex must be negative, i.e., $(2\ga-1)(-4\ga^2+2\ga+1)<0$.
Since $\ga > 1/\sqrt{2}$ implies that $2\ga-1 > 0$, the two conditions are jointly equivalent to the single inequality $4\ga^2-2\ga-1>0$. This final inequality holds if and only if $\ga > (1+\sqrt{5})/4$.
\qed

\begin{claim}
$\mathcal{G}(\ga)\subseteq \mathcal{G}(\ga')$ for $\ga < \ga'$.
\end{claim}
\noindent
This claim trivially holds if $\mathcal{G}(\ga)$ is empty. 
  Suppose $\mathcal{G}(\ga)$ is nonempty. 
  Then, $\alpha<\beta$ must hold; otherwise, $f(\alpha,\beta,\gamma)\geq 1-\alpha (2\ga\beta-1)>0$.
Differentiating $f(\alpha,\beta,\gamma)$ with respect to $\gamma$ gives
$$\frac{\partial f(\alpha,\beta,\gamma)}{\partial \gamma}=2(\alpha-\beta)+2\ga(\alpha-\beta)^2-2\alpha\beta=-2(\beta-\alpha)(1-\ga (\beta-\alpha))-2\alpha \beta<0.$$
Thus, $f(\alpha,\beta,\gamma)$ is strictly decreasing in $\ga$ whenever $\alpha < \beta$. \qed

\subsection{Derivation of the unique linear equilibrium under $\mathcal{I}'$}\label{sectionI'}
Let Bob's and Carol's linear strategies be $a'_B=b_B^xx_B+b_B^yy$ and $a'_C=b_C^xx_C+b_C^zz$, respectively. 
Then, Bob's first-order condition is
\begin{align*}
a'_B&= (1-\beta)(\ga x_B+(1-\ga)y)+\beta (b_C^x+b_C^z)(\ga x_B+(1-\ga)y)\\
&= \ga\left(1-\beta+\beta (b_C^x+b_C^z)\right)x_B+(1-\ga)\left(1-\beta+\beta (b_C^x+b_C^z)\right)y.
\end{align*}
By matching the coefficients, Bob's equilibrium slopes satisfy
\begin{align*}
b^x_B&= \ga(1-\beta+\beta (b_C^x+b_C^z));\\
b^y_B&= (1-\ga)(1-\beta+\beta (b_C^x+b_C^z)).
\end{align*}
A similar logic applies to Carol's equilibrium strategy, and her equilibrium slopes satisfy
\begin{align*}
b^x_C&= \ga\left(1-\beta+\beta (b_B^x+b_B^y)\right);\\
b^z_C&= (1-\ga)\left(1-\beta+\beta (b_B^x+b_B^y)\right).
\end{align*}
Solving these simultaneous equations yields $b_B^y=b_C^z=1-\ga$ and $b_B^x=b_C^x=\ga$.
Substituting the obtained Bob's equilibrium strategy into Ann's first-order condition gives
\begin{align*}
a'_A&=(1-\alpha)(\ga x_A+(1-\ga)y)+\alpha \left((1-\ga)y+\ga (\ga x_A+(1-\ga)y)\right)\\
&=(1-\ga)(1+\ga\alpha)y+ (\ga(1-\alpha)+\ga^2\alpha)x_A.
\end{align*}
This concludes the proof.
\qed

\subsection{Derivation of inequality (\ref{equationABC2})}
Let $U'_A(\tx, \ty, G)$ denote Ann's equilibrium payoff under $\mathcal{I}'$.
By the proof in Section \ref{sectionI'}, Ann's and Bob's equilibrium strategies under $\mathcal{I}'$ are given by
$a'_{A}= (1-\ga)(1+\ga \alpha)y+ \ga(1-\alpha + \ga\alpha) x_A$ and $a'_{B}=(1-\ga)y+\ga x_B$, respectively.
Notice that $a'_A$ and $a'_B$ coincide with the equilibrium strategies of agents with centralities of $1+\ga \alpha$ and $1$ under $\mathcal{I}$, respectively.
Thus, plugging  $1+\ga \alpha$ as Ann's centrality and $1$ as Bob's centrality into (\ref{equationpayoff}) yields
\begin{align*}
U'_A(\tx, \ty, G)&=  \tx^{-1}(1-\ga)((1+\ga \alpha)^2+\alpha)-\tx^{-1}(1+\alpha).
\end{align*}
The difference in Ann's equilibrium payoff between $\mathcal{I}'$ and $\mathcal{I}$ is then computed as
\begin{align*}
U'_A(\tx, \ty, G)-U_A(\tx, \ty, G)&=  \tx^{-1}(1-\ga)\left((1+\ga \alpha)^2+\alpha-(1-\ga\beta)^{-2}f(\alpha,\beta,\gamma)\right).
\end{align*}
Thus, $U_A(\tx, \ty, G)<U'_A(\tx, \ty, G)$ if and only if $f(\alpha,\beta,\gamma)<(1-\ga\beta)^2\left((1+\ga \alpha)^2+\alpha\right)$.
\qed

\subsection{Proposition \ref{propwelfare}}
By the proof of Proposition \ref{propindividual}, for each $i \in N$,
\begin{align*}
		\Delta U_i(\tx, \ty, G) &=  \tx^{-1}(1-\ga)\left(c_i^2(\ga, G)-\sum_{j \neq i}g_{ij}c_j(\ga, G)\left(c_j(\ga, G)-2\right)\right).
\end{align*}
Summing these equations over $i$ yields
\begin{align*}
	\Delta W(\tx, \ty, G) &=  \sum_{i\in N}\Delta U_i(\tx, \ty, G)\\
	&=\tx^{-1}(1-\ga)\left(\sum_{i\in N}c_i^2(\ga, G)-\sum_{i\in N}\sum_{j \neq i}g_{ij}c_j(\ga, G)\left(c_j(\ga, G)-2\right)\right)\\
	&=\tx^{-1}(1-\ga)\left(\sum_{i\in N}c_i^2(\ga, G)-\sum_{i\in N}d_i^{in}c_i(\ga, G)\left(c_i(\ga, G)-2\right)\right)\\
	&=\tx^{-1}(1-\ga)\sum_{i\in N}c_i(\ga, G)\left((1-d_i^{in})c_i(\ga, G)+2d_i^{in}\right).
\end{align*}
Hence, $\Delta W(\tx, \ty, G)<0$ if and only if $\sum_{i\in N}c_i(\ga, G)\left((1-d_i^{in})c_i(\ga, G)+2d_i^{in}\right)<0$. \qed

\subsection{Claims in Example \ref{exampleWelfare}}\label{sectionproofexample2}
\begin{claim}
Suppose $G$ is a core-periphery interaction network. Then, $\Delta W(\tx,\ty, G)<0$ if and only if  $f(\alpha,\beta,\gamma)<-l(1-\beta(2\ga\beta-1))/m$.
\end{claim}
\noindent 
Suppose $G$ is a core-periphery interaction network.
Each agent $i$'s  centrality is given by
$c_i(\ga, G)=1/(1-\ga \beta)$ if $i$ is core and $c_i(\ga, G)=(1+\ga (\alpha-\beta))/(1-\ga \beta)$ if $i$ is peripheral.
The sum of all core agents' in-degrees is $l\beta+m\alpha$, and the sum of all peripheral agents' in-degrees is $0$.
Plugging these into the left-hand side of (\ref{inequalitywelfare})  yields
\begin{align*} 
\mathrm{(LHS)} &=   (1-\ga\beta)^{-2}\left(l-(l\beta+m\alpha) +2(1-\ga\beta)(l\beta+m\alpha)+m(1+\ga (\alpha-\beta))^2\right)\\
 &=(1-\ga\beta)^{-2}\left(l+(l\beta+m\alpha)(1 -2\ga\beta)+mf(\alpha,\beta,\ga)+m\alpha(2\ga\beta-1)\right)\\
 &=(1-\ga\beta)^{-2}\left(mf(\alpha,\beta,\ga)+l(1-\beta(2\ga\beta-1)\right)).
\end{align*}
Hence, (\ref{inequalitywelfare}) holds if and only if $f(\alpha,\beta,\ga)<-l(1-\beta(2\ga\beta-1))/m$. \qed

\begin{claim}
$\mathcal{H}(\ga, q)\subseteq \mathcal{H}(\ga, p)$ for $p \leq q$. 
\end{claim}
\noindent This is immediate from the fact that $1-\beta(2\ga\beta-1)>0$.
\qed

\begin{claim}
For each $\ga\in (0, 1)$, $\mathcal{H}(\ga, l/m)\rightarrow \mathcal{G}(\ga)$ as $l/m\rightarrow 0$. 
\end{claim}
\noindent Fixing $\ga\in (0, 1)$, the right-hand side of (\ref{exampleinequalitywelfare}) converges to 0 as $l/m\rightarrow 0$. 
The obtained inequality in the limit coincides with inequality (\ref{equationABC}).
\qed

\subsection{Corollary \ref{propmonotonicity}}
Suppose $n \geq 3$. I construct two interaction networks $G$ and $G'$ such that $G \leq G'$ and $\Delta W(\tau_x, \tau_y, G') < 0 < \Delta W(\tau_x, \tau_y, G)$.
First, when $G$ is the empty network, defined by setting $\alpha = \beta = 0$ in the core-periphery structure of Example \ref{exampleWelfare}, $\Delta W(\tau_x, \tau_y, G) > 0$ holds for any $\gamma <1$.
Second, by the derivation in Example \ref{exampleWelfare}, the welfare effect is negative if the interaction parameters $(\alpha, \beta)$ belong to the set $\mathcal{H}(\gamma, l/m)$, defined by
\begin{equation*}
f(\alpha, \beta, \gamma) < -\frac{l}{m}(1-\beta(2\gamma\beta-1)).
\end{equation*}
Consider the limit as $\gamma \to 1$ and $\beta \to 1$ with fixed $\alpha = 1/2$. The left-hand side converges to $(1 + (1/2 - 1))^2 - 1/2 = -1/4$, while the right-hand side converges to $0$. By continuity, there exists a sufficiently high $\gamma < 1$ and parameters $(\alpha', \beta') \in (0, 1)^2$ such that the inequality holds. Let $G'$ be the network characterized by these parameters. Then, $\Delta W(\tau_x, \tau_y, G') < 0$.
Since $G$ has zero weights and $G'$ has strictly positive weights, $G \leq G'$. Thus, increasing connectivity from $G$ to $G'$ reverses the sign of the welfare effect.
\qed

\subsection{Derivation of the unique linear equilibrium under $\mathcal{I}^\dagger$}
\label{sectionproofeqmshare}
Following the proof in Section \ref{sectionproof1}, it is easy to show that the sum of equilibrium slopes must be one for each agent in any linear equilibrium under $\mathcal{I}^\dagger$.
This observation immediately implies that agent $i \neq 1$'s unique linear equilibrium strategy must be  $a^\dagger_i=x_i$.
Plugging these into agent $1$'s first-order condition yields
\begin{align*}
a^\dagger_1= (1-d_1^{out}) \mE_1[\theta]+\sum_{j \neq 1}g_{1j} \mE_1[x_j]=\mE_1[\theta]=\ga x_1+(1-\ga)y. 
\end{align*}
This concludes the proof. \qed

\subsection{Derivation of $U^\dagger_i(\tx,\ty, G)$ and $W^\dagger(\tx,\ty, G)$}
First, I derive agent $1$'s equilibrium payoff under $\mathcal{I} ^\dagger$:
\begin{align*}
	U^\dagger_1(\tx,\ty, G) &= -\left(1-d_1^{out}\right)\mE[(a^\dagger_1-\theta)^2] - \sum_{j \neq 1}g_{1j}\mE[(a^\dagger_1-a^\dagger_j)^2]\\
    &=-\mE[(a^\dagger_1-\theta)^2]- \sum_{j \neq 1}g_{1j}\mE[(a^\dagger_j-\theta)^2]+2 \sum_{j \neq 1}g_{1j}\mE[(a^\dagger_1-\theta)(a^\dagger_j-\theta)]\\
    &=-\mE[(a^\dagger_1-\theta)^2]- \sum_{j \neq 1}g_{1j}\mE[(a^\dagger_j-\theta)^2]\\
    &= -(1-\ga)^2\ty^{-1}-\ga^2 \tx^{-1}-d_1^{out}\tx^{-1}\\
    &= -\ga \tx^{-1}-d_1^{out}\tx^{-1}\\
    &= \tx^{-1}(1-\ga)-\tx^{-1}(1+d_1^{out}).
\end{align*}
Similarly, agent $i \neq 1$'s equilibrium payoff under $\mathcal{I} ^\dagger$ is derived as
\begin{align*}
	U^\dagger_i(\tx,\ty, G) &= -\tx^{-1}- \sum_{j \neq i, 1}g_{ij} \tx^{-1}-g_{i1}((1-\ga)^2\ty^{-1}+\ga^2 \tx^{-1})\\
    &= \tx^{-1}(1-\ga)g_{i1}-\tx^{-1}(1+d_i^{out}).
\end{align*}
Thus, $U^\dagger_1(\tx,\ty, G)= \tx^{-1}(1-\ga)+U_1(\tx, 0, G)$ and $U^\dagger_i(\tx,\ty, G)=\tx^{-1}(1-\ga)g_{i1}+U_i(\tx, 0, G)$ for each $ i \neq 1$.
Summing over all agents then yields $W^\dagger(\tx,\ty, G)=\tx^{-1}(1-\ga)(1+d^{in}_{1})+W(\tx, 0, G)$, which concludes the proof. \qed

\subsection{Proposition \ref{propinfoshare}}
By the proof of Proposition \ref{propindividual} and Proposition \ref{propwelfare}, 
\begin{align*}
	U^\dagger_1(\tx,\ty, G)-U_1(\tx,\ty, G)&=\tx^{-1}(1-\ga)+U_1(\tx, 0, G)-U_1(\tx,\ty, G)\\
	&=\tx^{-1}(1-\ga)-\Delta U_1(\tx,\ty, G)\\
	&=  \tx^{-1}(1-\ga)\left(1-c_1^2(\ga, G)+\sum_{j \neq 1}g_{1j}c_j(\ga, G)\left(c_j(\ga, G)-2\right)\right);\\
	W(\tx,\ty, G)-W^\dagger(\tx,\ty, G)&=W(\tx,\ty, G)-\tx^{-1}(1-\ga)(1+d^{in}_{1})-W(\tx, 0, G)\\
	&=\Delta W(\tx,\ty, G)-\tx^{-1}(1-\ga)(1+d^{in}_{1})\\
    &=\tx^{-1}(1-\ga)\left(\sum_{i\in N}c_i(\ga, G)\left((1-d_i^{in})c_i(\ga, G)+2d_i^{in}\right)-(1+d^{in}_{1})\right).
\end{align*}
Thus, $U_1(\tx,\ty, G)<U^\dagger_1(\tx,\ty, G)$ and $W^\dagger(\tx,\ty, G)<W(\tx,\ty, G)$ if and only if $c_1^2(\ga, G)-\sum_{j \neq 1}g_{1j}c_j(\ga, G)\left(c_j(\ga, G)-2\right)<1$ and $1+d^{in}_{1}<\sum_{i\in N}c_i(\ga, G)\left((1-d_i^{in})c_i(\ga, G)+2d_i^{in}\right)$. \qed

\subsection{Derivation of compound inequality (\ref{exampleinequalityinfoshare})}
\noindent
Suppose $G$ is a core-periphery interaction network. By substituting the simplified expressions for the left-hand side of (\ref{inequalityinfoshare1}) (derived in Section \ref{sectionproofexample1}) and the right-hand side of (\ref{inequalityinfoshare2}) (derived in Section \ref{sectionproofexample2}), the two conditions can be combined into the single compound inequality:
$$(1-\ga\beta)^{-2}f(\alpha,\beta,\gamma) < 1 < (1-\ga\beta)^{-2}\left(mf(\alpha,\beta,\gamma)+l(1-\beta(2\ga\beta-1))\right).$$
Multiplying all parts of this inequality by $(1-\ga\beta)^{2}$ yields the compound inequality (\ref{exampleinequalityinfoshare}). \qed

\subsection{Proposition \ref{prop:marginal_welfare}}
Differentiating $W(\tx, \ty, G)$ with respect to $\ty$ yields
\begin{equation*}
\frac{\partial W}{\partial \ty} = \tx^{-1} \left[ \frac{\partial (1-\ga)}{\partial \ty} S(\ga) + (1-\ga) \frac{dS(\ga)}{d\ga} \frac{\partial \ga}{\partial \ty} \right].
\end{equation*}
Using the definition $\ga = \tx/(\tx+\ty)$, the partial derivatives are calculated as
\begin{align*}
\frac{\partial \ga}{\partial \ty} &= -\frac{\tx}{(\tx+\ty)^2} = -\frac{\ga}{\tx+\ty}; \\
\frac{\partial (1-\ga)}{\partial \ty} &= \frac{\tx}{(\tx+\ty)^2} = \frac{\ga}{\tx+\ty}.
\end{align*}
Substituting these expressions into the derivative of welfare gives
\begin{align*}
\frac{\partial W}{\partial \ty} &= \tx^{-1} \left[ \frac{\ga}{\tx+\ty} S(\ga) - (1-\ga) \frac{\ga}{\tx+\ty} \frac{dS(\ga)}{d\ga} \right] \\
&= \frac{\ga}{\tx(\tx+\ty)} \left[ S(\ga) - (1-\ga) \frac{dS(\ga)}{d\ga} \right].
\end{align*}
Since $\ga, \tx, \ty > 0$,  $\partial W/\partial \ty < 0$ if and only if
\begin{equation}
S(\ga) < (1-\ga) \frac{dS(\ga)}{d\ga}. \label{eq:marginal_condition_step}
\end{equation}
Next, I derive the expression for $dS(\ga)/d\ga$. Differentiating $S(\ga)$ yields
\begin{align*}
\frac{dS(\ga)}{d\ga} &= \sum_{i \in N} \left[ \frac{\partial c_i}{\partial \ga} \left( (1-d_i^{in})c_i + 2d_i^{in} \right) + c_i \left( (1-d_i^{in})\frac{\partial c_i}{\partial \ga} \right) \right] \\
&= \sum_{i \in N} \frac{\partial c_i}{\partial \ga} \left( (1-d_i^{in})c_i + 2d_i^{in} + (1-d_i^{in})c_i \right) \\
&= 2 \sum_{i \in N} \frac{\partial c_i}{\partial \ga} \left( (1-d_i^{in})c_i + d_i^{in} \right).
\end{align*}
Using the identity $\partial \mathbf{c}/\partial \ga = \ga^{-1}(\mathbf{c}^\mathbf{c} - \mathbf{c})$, the scalar derivative for agent $i$ is $\partial c_i/\partial \ga = \ga^{-1}(c_i^\mathbf{c} - c_i)$. Substituting this into the expression for $dS(\ga)/d\ga$:
\begin{align*}
\frac{dS(\ga)}{d\ga} &= 2 \sum_{i \in N} \ga^{-1}(c_i^\mathbf{c} - c_i) \left( (1-d_i^{in})c_i + d_i^{in} \right) \\
&=  S'(\ga),
\end{align*}
which completes the proof.
\qed

\subsection{Proposition \ref{propwelfare_alt}}
First, I derive the equilibrium condition.
Differentiating the payoff function (\ref{alternative_payoff}) with respect to $a_i$ and setting the expected marginal utility to zero yields
\begin{equation*}
-(1-r_i)(a_i - \mE_i[\theta]) - r_i \sum_{j \neq i} g_{ij} (a_i - \mE_i[a_j]) = 0.
\end{equation*}
Rearranging terms, I obtain
\begin{equation*}
\left( 1 - r_i + r_i \sum_{j \neq i} g_{ij} \right) a_i = (1-r_i)\mE_i[\theta] + \sum_{j \neq i} r_i g_{ij} \mE_i[a_j].
\end{equation*}
Let $K_i \equiv 1-r_i + r_i d_i^{out}$. Dividing both sides by $K_i$ leads to
\begin{equation*}
a_i = \frac{1-r_i}{K_i} \mE_i[\theta] + \sum_{j \neq i} \frac{r_i g_{ij}}{K_i} \mE_i[a_j].
\end{equation*}
Defining $\tilde{g}_{ij} \equiv r_i g_{ij}/K_i$ and noting that $1 - \sum_{j \neq i} \tilde{g}_{ij} = (K_i - r_i d_i^{out})/K_i = (1-r_i)/K_i$, this equation simplifies to the first-order condition (\ref{FOC_alt}).
Since this condition is identical to that of the baseline model (\ref{FOC}) with $\tilde{G}$ replacing $G$, the unique linear equilibrium is given by $a_i^* = (1-\gamma)c_i(\gamma, \tilde{G})y + (1-(1-\gamma)c_i(\gamma, \tilde{G}))x_i$.

Next, following the derivation in the baseline model, expanding agent $i$'s payoff function (\ref{alternative_payoff}) yields
\begin{align*}
U_i (\tx,\ty, G)&= -(1-r_i)\mE[X_i^2] - r_i \sum_{j \neq i} g_{ij} \mE[X_i^2 - 2X_i X_j + X_j^2] \\
&= -K_i \mE[X_i^2] - \sum_{j \neq i} r_i g_{ij} \mE[X_j^2] + 2 K_i \sum_{j \neq i} \tilde{g}_{ij} \mE[X_i X_j],
\end{align*}
where the second equality follows from the definitions of $K_i$ and $\tilde{g}_{ij}$.
Subtracting the payoff in the absence of public information, $U_i(\tx, 0, G) = -\tx^{-1}(K_i+r_id_i^{out})$, and substituting the variance and covariance equilibrium moments derived in (\ref{identity1}), (\ref{identity2}), and (\ref{identity3}) yields the change in agent $i$'s payoff:
\begin{align*}
\Delta U_i (\tx,\ty, G) &= \tx^{-1}(1-\ga) \left( -K_i c_i(c_i-2) - \sum_{j \neq i} r_i g_{ij} c_j(c_j-2) + 2 K_i c_i(c_i-1) \right) \\
&= \tx^{-1}(1-\ga) \left( K_i c_i^2 - \sum_{j \neq i} r_i g_{ij} c_j(c_j-2) \right),
\end{align*}
where $c_i$ denotes $c_i(\ga, \tilde{G})$.
Summing over all agents to obtain $\Delta W = \sum_{i \in N} \Delta U_i$ and exchanging indices in the double summation term:
\begin{align*}
\sum_{i \in N} \sum_{j \neq i} r_i g_{ij} c_j(c_j-2) &= \sum_{j \in N} c_j(c_j-2) \sum_{i \neq j} r_i g_{ij} \\
&= \sum_{j \in N} d_j^{in, r} c_j(c_j-2),
\end{align*}
where $d_j^{in, r}=\sum_{i \neq j} r_i g_{ij} $.
Thus, aggregate welfare decreases if and only if
\begin{equation*}
\sum_{i \in N} \left( K_i c_i^2 - d_i^{in, r} c_i(c_i-2) \right) < 0,
\end{equation*}
as desired. \qed

\subsection{Proof of Corollary \ref{prop:dual_heterogeneity}}
The welfare effect is positive if and only if $\sum_{i \in N} \Omega_i > 0$, where $\Omega_i = c_i [ (K_i - d_i^{in, r})c_i + 2d_i^{in, r} ]$.
I show that this sum is positive in both cases.
\paragraph{Case 1: Homogeneous Intensity ($r_i = r$ and $d_i^{out} = 1$).}
In this case, $K_i = 1$ and $d_i^{in, r} = r d_i^{in}$. The normalized matrix $\tilde{G}$ has elements $\tilde{g}_{ij} = r g_{ij}$, and its row sums are uniform as $\tilde{d}^{out}_i = r$. The centrality is also uniform across agents, given by $c_i (\ga, \tilde{G})= c = (1-\gamma r)^{-1}$.
Factoring out the constant $c$, I obtain
\begin{equation*}
\sum_{i \in N} \Omega_i = c^2 \sum_{i \in N} (1 - r d_i^{in}) + 2cr \sum_{i \in N} d_i^{in}.
\end{equation*}
Since $\sum d_i^{in} = \sum d_i^{out} = n$, this simplifies to $n c^2 (1-r) + 2crn$, which is strictly positive for $r < 1$.
\paragraph{Case 2: Homogeneous Direction ($g_{ij} = \frac{1}{n-1}$).}
 The total payoff weight in this case is $K_i = 1 - r_i + r_i = 1$ since $d_i^{out} = 1$ by assumption.
The effective in-degree is the average intensity of neighbors:
\begin{equation*}
d_i^{in, r} = \sum_{j \neq i} r_j g_{ji} = \frac{1}{n-1} \sum_{j \neq i} r_j.
\end{equation*}
It follows that $d_i^{in, r} < 1$ for all $i$ since $r_j < 1$ for all $j \neq i$ by assumption.
The coefficient of the squared centrality term in $\Omega_i$ is $K_i - d_i^{in, r} = 1 - d_i^{in, r} > 0$.
Thus, every individual welfare contribution $\Omega_i$ is strictly positive, which completes the proof.
\qed

\subsection{Proposition \ref{prop:efficient_use}}
First, I derive the first-order condition for the efficient strategy.
The planner maximizes aggregate welfare $w(a, \theta) \equiv \sum_{i \in N} u_i(a, \theta)$ subject to the constraint that each agent $i$'s action $a_i$ depends only on their local information set $\mathbf{s}_i$.
Differentiating $w$ with respect to agent $i$'s strategy $a_i$ and setting the expected marginal welfare to zero yields
\begin{equation*}
-\left(1 - \sum_{j \neq i} g_{ij}\right) (a_i - \mathbb{E}_i[\theta]) - \sum_{j \neq i} g_{ij} (a_i - \mathbb{E}_i[a_j]) - \sum_{j \neq i} g_{ji} (a_i - \mathbb{E}_i[a_j]) = 0.
\end{equation*}
The first two terms represent agent $i$'s private marginal utility, while the third term captures the externality $i$'s action imposes on neighbors $j$ who seek to coordinate with $i$.
Rearranging terms, I obtain
\begin{equation*}
\left( 1 - \sum_{j \neq i} g_{ij} + \sum_{j \neq i} g_{ij} + \sum_{j \neq i} g_{ji} \right) a_i = \left(1 - \sum_{j \neq i} g_{ij}\right) \mathbb{E}_i[\theta] + \sum_{j \neq i} (g_{ij} + g_{ji}) \mathbb{E}_i[a_j].
\end{equation*}
Simplifying the coefficient on the left-hand side to $1 + d_i^{in}$ and dividing both sides by this term leads to
\begin{equation*}
a_i = \frac{1 - d_i^{out}}{1 + d_i^{in}} \mathbb{E}_i[\theta] + \sum_{j \neq i} \frac{g_{ij} + g_{ji}}{1 + d_i^{in}} \mathbb{E}_i[a_j].
\end{equation*}
Defining the fictitious interaction weights $g_{ij}^F \equiv (g_{ij} + g_{ji})/(1 + d_i^{in})$ and noting that
\begin{equation*}
1 - \sum_{j \neq i} g_{ij}^F = 1 - \frac{\sum_{j \neq i} g_{ij} + \sum_{j \neq i} g_{ji}}{1 + d_i^{in}} = \frac{1 + d_i^{in} - d_i^{out} - d_i^{in}}{1 + d_i^{in}} = \frac{1 - d_i^{out}}{1 + d_i^{in}},
\end{equation*}
this equation simplifies to the desired first-order condition (\ref{efficient_FOC}).

To ensure the existence of a unique linear efficient strategy, I verify that the spectral radius of $G^F$ satisfies $\gamma \rho(G^F) < 1$.
Since $G^F$ is a non-negative matrix, its spectral radius is bounded by its maximum row sum.
Using the assumption $\sum_{j \neq i} g_{ij} < 1$, the row sum for any agent $i$ is strictly less than one:
\begin{equation*}
\sum_{j \neq i} g_{ij}^F = \frac{d_i^{out} + d_i^{in}}{1 + d_i^{in}} = 1 - \frac{1 - d_i^{out}}{1 + d_i^{in}} < 1.
\end{equation*}
Thus, $G^F$ is row-substochastic, ensuring that the inverse $(I - \gamma G^F)^{-1}$ exists and is non-negative.
Since the condition (\ref{efficient_FOC}) is identical to the equilibrium best-response condition (\ref{FOC}) with $G^F$ replacing $G$, the unique linear efficient strategy is given by the unique linear equilibrium of the game defined by $G^F$.
Applying Proposition \ref{propequilibrium} with $G^F$, agent $i$'s efficient strategy is thus
\begin{equation*}
a_i^{F} = (1-\gamma)c_i(\gamma, G^F) y + (1 - (1-\gamma)c_i(\gamma, G^F)) x_i,
\end{equation*}
where $c_i(\gamma, G^F)$ is the Katz-Bonacich centrality in the fictitious network $G^F$.
\qed

\newpage
\bibliography{hetpublic} 

\end{document}